\documentclass[preprint]{aastex}

\usepackage{graphicx}

\slugcomment{\today}

\shorttitle{Time spectroscopy of AW~UMa}
\shortauthors{Rucinski}

\begin{document}

\title{Time sequence spectroscopy of AW~UMa.\\
The 518 nm Mg~I triplet region analyzed with
Broadening Functions\footnote{Based on observations 
obtained at the Canada-France-Hawaii Telescope (CFHT) which 
is operated by the National Research Council of Canada, 
the Institut National des Sciences de l'Univers of the Centre 
National de la Recherche Scientique of France, 
and the University of Hawaii.}}

\author{Slavek M. Rucinski} 
\affil{Department of Astronomy and Astrophysics, 
University of Toronto \\
50 St. George St., Toronto, Ontario, M5S~3H4, Canada}
\email{rucinski@astro.utoronto.ca}

\begin{abstract}
High resolution spectroscopic observations of AW~UMa, obtained on 
three consecutive nights with the median time resolution of 2.1 minutes,
have been analyzed using the Broadening Functions method in the
spectral window of 22.75 nm around the 518 nm Mg~I triplet region.
Doppler images of the system reveal the presence of vigorous
mass motions within the binary system; their presence puts into question 
the solid-body rotation assumption of the contact binary model. 
AW~UMa appears to be a very tight, semi-detached binary; 
the mass transfer takes place from the more massive to the 
less massive component. The primary, a fast-rotating star with 
$V \sin i = 181.4 \pm 2.5$ km~s$^{-1}$, is covered by inhomogeneities: 
very slowly drifting spots and a dense network of ripples more 
closely participating in its rotation. The spectral lines
of the primary show an additional broadening component 
(called the ``pedestal'') which originates either in the
equatorial regions which rotate faster than the rest of 
the star by about 50 km~s$^{-1}$ or in an external disk-like
structure. The secondary component appears to be smaller than 
predicted by the contact model. The radial velocity field around 
the secondary is dominated by accretion of matter transferred 
from (and possibly partly returned to) the primary component. 
The parameters of the binary are:
$A \sin i = 2.73 \pm 0.11\, R_\odot$ and 
$M_1 \sin^3 i = 1.29 \pm 0.15\, M_\odot$, 
$M_2 \sin^3 i = 0.128 \pm 0.016\, M_\odot$.
The mass ratio $q_{\rm sp} = M_2/M_1 = 0.099 \pm 0.003$, while 
still the most uncertain among the spectroscopic elements, is 
substantially different from the previous numerous and mutually
consistent photometric investigations which were based on 
the contact model. It should be studied why photometry and
spectroscopy give so very discrepant results and whether AW~UMa is
an unusual object or that only very high-quality spectroscopy 
can reveal the true nature of W~UMa-type binaries.
\end{abstract}
    

\section{Introduction}
\label{intro}

AW~UMa (HD~99946) is a well known, very close, short-period 
($P = 0.4387$ d) binary system  belonging to the class 
of ``W~UMa-type'' binary stars.
It is sometimes called ``Paczynski's star'' for the discoverer
\citep{BeP1964}, who recognized its particular importance 
for understanding W~UMa-type binaries. This is due to its shallow, 
total, long-duration eclipses with equal depth, the property
which directly indicates that the two very different stars 
have the same surface temperature. 
The first light curve synthesis models 
\citep{MD1972,Wilson1973} showed that photometric variability of 
AW~UMa is in perfect agreement with the model for W~UMa binaries
presented by \citet{Lucy1968a,Lucy1968b}. The Lucy model is based
on the assumption of a solid-body rotation which permits the definition 
of equipotentials; one equipotential is common to both stars. 
In the case of AW~UMa, the agreement of observed light curves
with model predictions holds particularly well 
in spite of the large disparity in masses of the components, as
estimated from the Lucy model. 
Several photometric investigations, continuing to the recent 
study of \citet{Wilson2008}, confirmed that the light curve 
of AW~UMa is in full agreement with the one predicted by the Lucy 
model for the very small mass-ratio,  
$q_{\rm ph} = M_2/M_1 \simeq 0.07 - 0.08$. 
The perfect agreement with the model was reassuring 
but somewhat unexpected because the model is built on the rather
restrictive assumption of the strictly solid-body rotation, i.e.\ 
absence of velocity field in the rotating system of coordinates. 
This would also imply a total absence of any {\it differential\/} 
rotation for the primary F2-type star. In addition, 
because of the disparate masses, we know that the stars 
forming AW~UMa must be very different in terms of their internal 
structure, with the more massive primary component 
($M_1 \simeq 1.3 \,M_\odot$) providing all the
energy to be radiated by the whole system; the 10 -- 12 times 
less massive secondary component carries the angular momentum 
of the system and -- within the framework of Lucy's model --
is probably an isothermal structure.
Yet the common envelope seemed to be 
somehow oblivious to these structural differences. Indeed,
\citet{BeP2007} suggested, following \citet{Step2006},
that the binary consists of a Main Sequence star and a helium star,
a core of the former primary of AW~UMa, currently submerged in
the material of the primary component.

While an abundance of photometric light-curve-synthesis solutions
confirmed the applicability of Lucy's model to AW~UMa, 
the radial velocity (RV) data were, for long time,
discordant and inconclusive  
\citep{McLean1981,Rens1985,Rci1992,Prib1999}. 
In general they did not contradict the photometric results, but
neither did they contribute much new information because of the poor 
definition of the radial-velocity orbital amplitudes:
On one hand, the secondary was difficult to detect, while the
minuscule orbital motion of the primary was hard to measure accurately
because of its very broad spectral lines. 
In contrast to photometric observations, spectroscopic analyses 
require large telescopes to insure good signal-to-noise values
at short exposures times. The study of radial velocities conducted at the 
David Dunlap Observatory (DDO) using the 1.9m telescope 
\citep{PR2008} was a breakthrough: One can see the improvement
by comparing Fig.~2 in \citet{Wilson2008} to Fig.~6 in \citet{PR2008}.
It was found that the orbital velocity amplitudes
clearly indicate a small mass ratio, $q_{\rm sp} = 0.10$, 
which is however substantially different from $q_{\rm ph}$ 
by several formal errors of the photometric solutions. From the 
spectroscopic observations, the secondary component did 
not seem to look like a star, but perhaps an accretion disk. 
Furthermore, the primary component showed surface  
features and an unexplained ``pedestal'' of large rotational 
velocities surrounding the rotationally broadened
profile. It appeared that AW~UMa {\it is not
a contact binary in the sense as defined by Lucy's model, but rather
a semi-detached binary; thus, the spectroscopic observations
strongly contradicted the photometric solutions which entirely
depend on the adoption of this model.} The results were so unexpected that
they were taken with disbelief or ignored. Suggestions were made 
(R.\ E.\ Wilson -- private communication) that the spectral
analysis using Broadening Functions (BF) had some problems which would need an
explanation before questioning the well established photometric results. 
This was unfortunate because (1)~one cannot hope to obtain any
spectral information from incredibly blended spectral lines without
some sort of a deconvolution method, (2)~the BF approach 
shares strengths and weakness with other deconvolution
methods utilizing simultaneously many spectral lines 
to extract velocity information, (3)~as such,
it provides direct image of the binary in the radial velocity space,
but the convenient normalization of the resulting BF's provides
a way to verify the assumed spectral type and to spectroscopically
estimate the metal content \citep{Rci2013a}. 
Besides, the BF method served very well for the whole DDO program of
radial velocities for bright close binaries 
(summaries in \citet{ruc2010,ruc2012}) and most of W~UMa 
binaries did appear to conform to the Lucy model. But none was so
thoroughly analyzed as the brightest observed AW~UMa so it is
not excluded that their agreement with the contact model could also
fail when subject to such scrutiny as is the case of that binary. 

The DDO observations of AW~UMa by \citet{PR2008} did have weaknesses related
to the data acquisition: The 109 spectra were obtained on 12 nights in
varying weather conditions, having exposure times ranging between 
5.5 and 15 minutes and with a moderate spectral resolution of 
about 20 km~s$^{-1}$ ($R \simeq 15,000$).
% within the spectral window with a width of 24 nm.
Observations described in the current paper improve on all of 
the above numbers by observing AW~UMa 
continuously on three consecutive nights using high resolution spectra.
The CFHT spectrograph had several times higher throughput and
efficiency while the telescope was two times larger than that used at DDO.  
The available spectra cover practically the whole optical region. 
The current paper describes the data for the Mg~I window for a direct 
comparison with the DDO results. Further analyses of the same
spectra are planned for the future publications. 

Section~\ref{obs} and \ref{BFs} describe the observational
data and their processing using the Broadening Functions technique. 
Properties of the primary component of AW~UMa are discussed in 
Section~\ref{pri} and \ref{inhom} while properties of the
secondary component are discussed in Section~\ref{sec}.
Section~\ref{binary} summarizes properties of AW~UMa as a binary system.
Section~\ref{further} contains conclusions and a discussion of
potential future work.

\section{CFHT observations}
\label{obs}

The observations of AW~UMa  were obtained 
on the nights of 11 -- 13 March 2011 using the 
Canada-France-Hawaii Telescope (CFHT) and its 
Espadons spectrograph working in the spectro-polarimetric mode.
The nominal spectral resolving power was $R \simeq 68,000$. 
Four consecutive observations with
different positions of the polarization analyzer can be used
to determine the Stokes circular polarization 
parameter $V$. In total, 146 series of polarization measurements
were obtained; their preliminary analysis by Dr.\ J.-F.\ Donati 
(private communication)
indicated no circular polarization at a noise level as low as
0.01\% and thus no obvious signature of surface magnetic fields.
For that reason, we used all $146 \times 4 = 584$ 
individual observations of AW~UMa
leaving the polarization issues for a possible future detailed
re-analysis. The exposure times were 90 seconds
and the median spacing between observations 
was 125 seconds. Some breaks in the continuous monitoring 
took place for observations of four standard stars 
and for the queue-observing housekeeping (visible as gaps in the
figures below). 

The data were processed by the CFHT pipeline ``Libre-Esprit'' 
\citep{Donati1997}\footnote{http://www.cfht.hawaii.edu/Instruments/Spectroscopy/Espadons/Espadons\_esprit.html}. 
While the full spectra extended from 
370 to 1,050 nm, we use here only the data in  
the window around the Mg~I triplet, 506.05 -- 528.80 nm, 
with three echelle orders contributing to this segment. This
choice was made to permit a direct comparison with the results 
of \citet{PR2008}.
The orbital phases were calculated following the discussion
of the photometric eclipse moments in \citet{Rci2013b} which was
based on the literature data merged with the recent 
{\it MOST\/} satellite results obtained soon after the CFHT
observations. The linear elements for the primary eclipse 
just before the start of the CFHT observations were 
$HJD - 2,400,000 = 55631.6498 + 0.43872420 \times E$,
where $E$ is the epoch. The nightly ranges of $HJD$ and phase 
(calculated from the above epoch)
are tabulated in Table~\ref{tab_obs}. Altogether, slightly more
than 5 orbital cycles of AW~UMa were observed.
% by 584 observations.   

\placetable{tab_obs}                  % Table 1   observing nights

\section{The Broadening Functions technique}
\label{BFs}

The Broadening Function (BF) technique was introduced for the study 
of AW~UMa some time ago \citep{Rci1992} and since then has been
considerably improved. It was extensively used during the DDO radial 
velocity studies of short-period binary stars between 1999 and 2008, 
which provided radial-velocity orbital data for 162 systems,
see \citet{ruc2010,ruc2012}. As with the current observations, 
the previous, 1992 study of AW~UMa was also based on the CFHT data, 
but used a relatively low throughput 
spectrograph--detector combination and did not not offer any 
substantial progress in our understanding of the binary. 

In essence, the BF technique is the determination of the Doppler
broadening kernel $B$ in the convolution equation,
$P(x) = \int B(x') T(x-x') dx'$, where $P$ is the spectrum of
the binary and $T$ is the spectral template. 
$P$ and $T$ are of the length of the adopted window. 
The window used here, 506.05 -- 528.80 nm corresponds to the span of 
$\simeq 13,200$ km~s$^{-1}$ in radial velocities.
The lengths of the Broadening Function $B$ is 
selected on one hand to be much shorter than 
$P$ and $T$ to provide adequate over-determinacy, yet long enough to
adequately cover the whole range of $B$ with a baseline on both
sides of the broadened profile.
In the current study, the adopted BF range was $[-495, +495]$ km~s$^{-1}$
permitting solution of the linear equations representing the convolution 
using the least-squares method with the 13-fold over-determinacy. 
The BF technique is very similar to the Least Squares Deconvolution (LSD)
method introduced by \citet{DC1997} and \citet{Donati1997} in that
it is a fully linear approach and thus far superior to
the Cross-Correlation Function (CCF). The BF technique normalizes the
results in such a way that the integral over all velocities, 
$I = \int B(v) dv$, is equal to unity for an exact spectral
match of the $P$ and $T$. This property and the dependence of
$I$ on the spectral type and metal abundance 
was used for spectroscopic estimates of the metallicities 
of W~UMa-type binaries in \citet{Rci2013a}.

During the DDO program, we used the actual spectra of 
slowly-rotating standard stars of matching spectral types
as templates $T$.
This permitted direct calibration of the zero point of the radial
velocities and assured inclusion of all lines in the spectral window.
Although spectral-standard stars were observed, 
the CFHT data did not permit this approach because the 
tiny deficiencies in the pipeline processing (in  
transformations from the 2D echelle images to 1D spectra, the same for
$P$ and $T$) became strongly amplified after the  array-inversion operation.
Instead, we used a model spectrum computed for the spectral type
of AW~UMa by Dr.\ Jano Budaj. The model spectrum covered the whole
optical spectrum for $T_{\rm eff} = 7250$~K,
$\log g = 4.5$ (cgs), with an assumed solar metallicity, 
a micro-turbulence of 2 km~s$^{-1}$ and no macro-turbulence. 

The CFHT pipeline-processed spectra were sampled at intervals
of 1.8 km~s$^{-1}$, but to insure the high quality of the BF's, 
they were smoothed with a Gaussian with a FWHM of 8.5 km~s$^{-1}$,
% 8.46 km~s$^{-1}$; 
which is equivalent to the spectral resolving power of
$R \simeq 35,000$. 
At this resolution, the input spectra have the high signal-to-noise
of $(S/N)_{\rm sp} \simeq 200 - 500$, 
depending on the orbital phase of the binary,
the instantaneous brightness of the star and on the
presence of cirrus or changes in the air mass; 
the median for all spectra is $(S/N)_{\rm sp} \simeq 345$. 
The BF determination involves inversion of very large arrays 
leading to a decrease in the signal-to-noise. For the current
choice of the BF length 
%(and thus the over-determinacy), 
$(S/N)_{\rm BF} \simeq 30 - 115$, with the median $(S/N)_{\rm BF} = 57$.
The BF errors are distributed randomly along the function and 
do not depend on its local value. 
The BF's for all observed spectra are available for retrieval, as 
in Table~\ref{tab_BFs}.

\begin{figure}%[ht]
\begin{center}
\includegraphics[width=11.5cm]{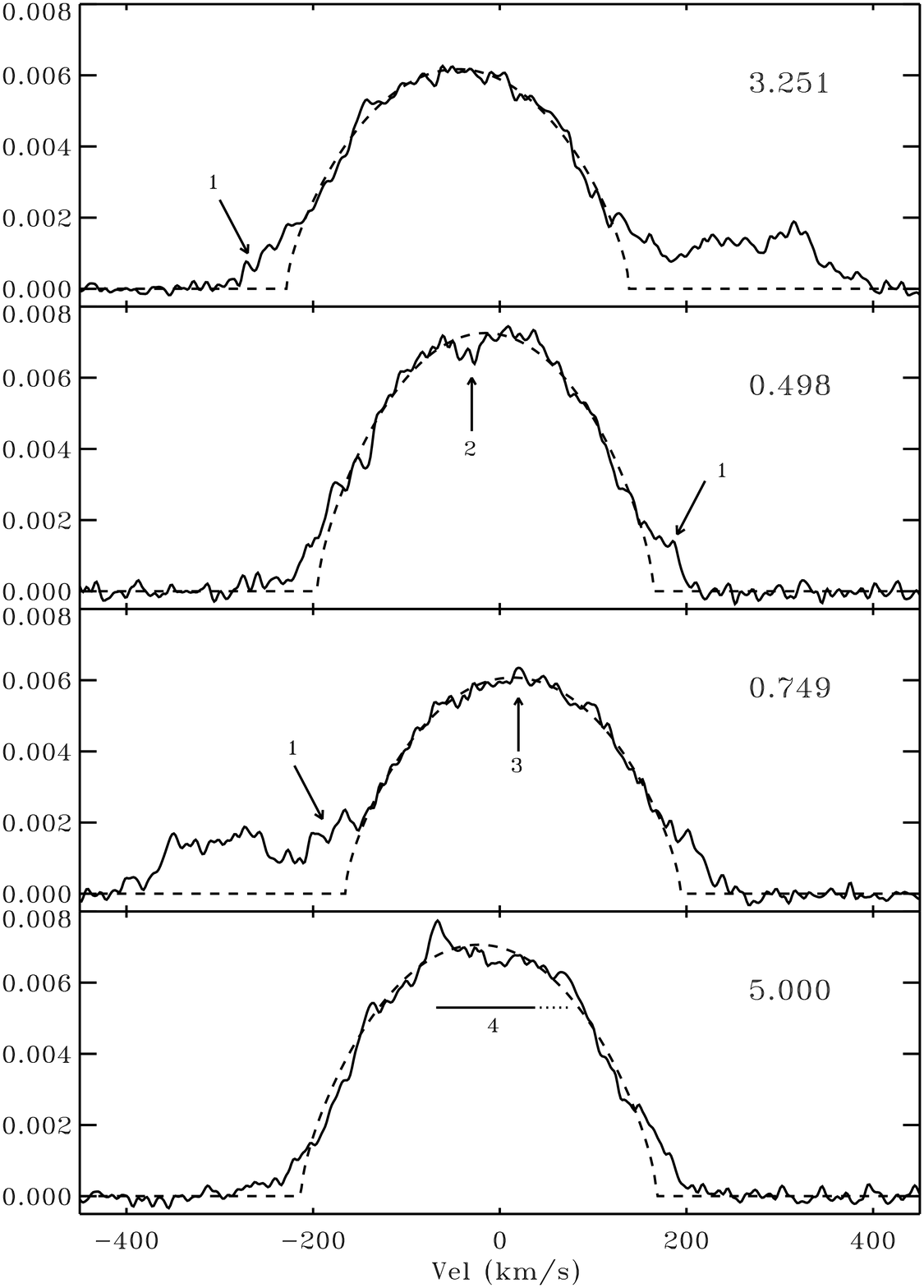}    
\caption{
\footnotesize 
Broadening Functions for four individual observations closest to
the orbital quadratures and the eclipse centers.
The phases -- counted from the primary eclipse before the start
of the observations -- are given in the figure panels. Velocities
in the horizontal axis are in the heliocentric system.
Units of the BF's in the vertical scale result from their 
normalization of the integral to unity for the exact spectral match 
of the template when sampled at 1.8 km~s$^{-1}$ (see the text). 
For AW~UMa, the integral is almost constant in phase
and close to 1.1 (see the lower panel of Figure~\ref{fig2}). 
The surface features on the primary component which are
discussed in detail in the paper are indicated by the 
numbers: 1 -- the pedestal, 2 -- a typical spot, 3 -- surface ripples,
4 -- the secondary component projecting onto the primary star
(see Section~\ref{sec}). 
The broken lines give the rotational profiles which were fit to 
the upper 45\%, 50\% and 55\% of the primary feature to obtain 
radial velocities of the primary component.
}
\label{fig1}
\end{center}
\end{figure}

\placefigure{fig1}                % fig.1   nights

\placetable{tab_BFs}              % Table 2     ASCII BFs

\begin{figure}%[ht]
\begin{center}
\includegraphics[width=11.5cm]{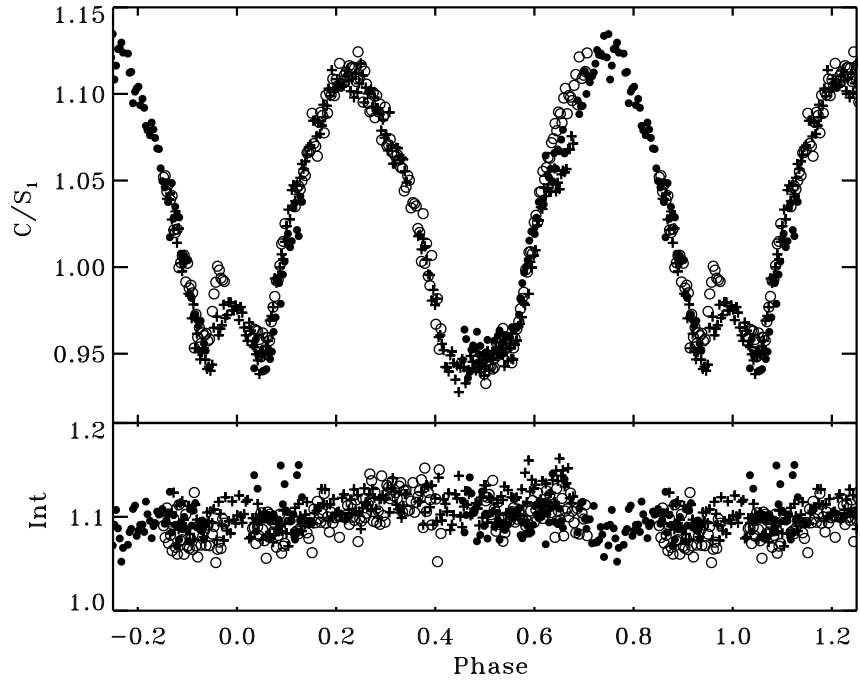}    
\caption{
\footnotesize 
The upper curve shows the values (shown as inverses) of the
central strength of the fitting rotational profile for
the primary component, $S_1$; they are normalized to the integral
of the rotational profile (the constant $C$, 
see the text). The symbols correspond to the three nights: 
night \#1, filled circles; night \#2, open circles; 
night \#3, crosses. 
The lower curve shows the integrals of the BF's.  
}
\label{fig2}
\end{center}
\end{figure}

Broadening Functions for four cardinal orbital phases  
are shown in Figure~\ref{fig1}. The BF 
technique produces functions which have integrals 
normalized to unity for the perfect spectral
match of the template spectrum to the observed spectrum in terms
of the overall strength; any deviations in the spectral type 
or metallicity manifest themselves through the integral 
being systematically smaller or larger than unity. 
For the adopted spectral model template of a F2V star, the integrals 
have the median value of 1.1; see the lower curve in Figure~\ref{fig2}. 
The deviations from unity may be interpreted 
that the spectral type of AW~UMa is slightly
later than the F2V model template. The 
corresponding shift in color would be then $\Delta(B-V) \simeq +0.03$ 
(see Figure~3 in \citet{Rci2013a}). The shift should
not necessarily be taken as a disagreement between the color of 
the star and the template used, because the spectral type of 
AW~UMa has never been determined through spectral classification.
Instead, it has always been evaluated by matching to the colors of
MS stars\footnote{This discrepancy
may also result from the systematic differences between intensity and 
flux spectra for strongly distorted stars which we disregard here.}. 
The small variation in the observed BF-integral
within 1.09 -- 1.12, with a slight increase over the half of the orbit
centered on the secondary eclipse (phases $0.25 < \phi < 0.75$)
may be interpreted as a tendency towards lower temperature
for the part of the orbit when the small star is 
behind the more massive star.
This does not agree with the standard contact model which 
predicts reddening to take place during both eclipses.

\placefigure{fig2}      % fig.2

The normalization property of the BF's to the integral of the template
offers a powerful check on their capability to capture the whole
light curve variations; it also permits one to ascertain 
that the phase variations observed spectroscopically correspond to
the same part of the binary which produces the photometric light variations. 
The check is based on fits of the rotational profile $r$
to the primary star feature, $R = S_1 \times r(V \sin i,u) + S_0$,
as described in Section~\ref{pri-fits} below. 
Here we use the normalization factor $S_1$ and relate its 
time variations to the integrated rotation profile, $C$, which is 
a constant. The ratio $C/S_1$ tells us how much light 
(i.e.\ which portion of the integral $I = \int B dv$)
comes from the whole binary system relative to 
the total light from the primary component. This leads to a rather well
defined ``light curve'' as in the upper panel of 
Figure~\ref{fig2}. This curve has been obtained 
entirely from the strength of the spectral lines,
without any reference to photometry. The amplitude of the
curve is about 16\% or 0.19 in magnitudes which is $\simeq$~3\%
less than observed photometrically in the $V$-band \citep{Prib1999}. 
The curve is not perfect as it shows apparent deterioration 
in rotational-profile fits during the primary eclipse 
when the secondary component transits the disk of the primary.
The inner contacts of ``totality'', as delineated by cusps in the curve 
at the primary eclipse (Figure~\ref{fig2}) are located at the phases
$-0.070$ and $+0.053$, each determined to $\pm 0.003$. The
less-well defined inner contact phases for the secondary eclipses,
when the secondary disappears behind the larger star,  are 
$-0.080$ and $+0.055$ ($\pm 0.005$). All these contacts are similar to
the well determined photometric contacts ($\pm 0.062$) which play
the essential role in the method of \citet{MD1972}. We return to the
eclipse contact phases when discussing the spectroscopic 
estimates of the secondary component size 
in Section~\ref{sec-size} and in Table~\ref{tab_contact}.

Thanks to the linear properties of the BF's,
plots such as in Figure~\ref{fig1} are expected to be 
faithful representations of the binary in the radial-velocity space. 
But this is true on condition of a strict correspondence of 
the photospheric star spectrum to the template spectrum. 
This condition must be obeyed by any technique that utilizes spectral
deconvolution, not only the BF but also the CCF and the LSD methods. 
Without these methods we could not analyze spectra equally complex
as those of AW~UMa, but there is a cost involved: 
The presence of emissions at any place on the star or of local
variations of temperature may distort the BF's and spoil
the radial-velocity mapping of the stellar surfaces. 
In fact, the spectral lines appearing fully in emission would 
produce a negative BF. Thus, when
the spectral lines are shallower than in the template,
then the BF will show depressions, but when the lines are
deeper, then the BF will be stronger. 
The lines can be deeper for surface regions of lower 
temperature while the lines can be shallower for higher local surface
temperatures or when the lines are filled in by emissions. 
For the spectral type of AW~UMa, the Fe~II lines dominate strongly 
over the Mg~I 518.4 nm triplet lines in our window 
\citep{Rci2013a}. The magnesium lines are not expected to 
show emission in this temperature regime, although we cannot 
entirely exclude this possibility. 
As we will see, our results reveal a very complex picture
of AW~UMa which is hard to explain through strictly 
geometric causes alone. Thus, 
it is possible that some depressions in the BF's are not due
to dark photospheric spots but, rather, 
reflect the presence of emissions (or -- correspondingly -- of
a shallower atmospheric source function than in the 
template model); conversely, segments in the BF's of unusually high
intensity may indicate lower local temperatures.

\section{The primary component} 
\label{pri}

\subsection{The rotational profile fits}
\label{pri-fits}

The spectral features of the primary component were fit for all 
BF's by rotational profile of a rigidly rotating, limb-darkened star:
$R = S_1 \times r(V \sin i,u) + S_0$, where $r(V \sin i,u)$
is the center-normalized profile (as given in many text-books,
e.g.\ \citet{gray2005}) and $S_i$ are the two fit constants. The 
profile $r$ involves two stellar parameters,
the projected equatorial velocity, $V \sin i$, which is basically
the scaling constant of the $X$-axis, and the
limb darkening coefficient, here assumed constant, $u = 0.6$. The profile   
centroid position, $V_1$, gives the mean velocity of the 
primary star and is the fourth free parameter of the fit.

The least-square fits were done using the upper 50\% of the profiles 
(within $-142 < V < +142$ km~s$^{-1}$ for the assumed mean 
$V \sin i = 181.4$ km~s$^{-1}$, see below in Section~\ref{vsini}), 
to avoid complications from the ``pedestal'' at large 
rotational velocities around the primary (Section~\ref{pedest}). 
Also, one can expect a lesser influence of the equatorial flows 
on the rotational profile for high latitudes of the primary component. 
The typical fits can be seen in Figure~\ref{fig1}. 

The rotational profile fits were well determined for 
all phases, as can be judged by the small scatter in the fit parameters.
The baseline level $S_0$ has a very small scatter around zero 
with the median error of $\pm 6 \times 10^{-6}$ 
in the integral BF units (as in Figure~\ref{fig1}); when 
expressed relative to the maximum strength (called below ``i.u.''), 
the median error of $S_0$ is only 0.01\%. 
The scaling factor $S_1$ shows systematic variation with phase, as
shown in Figure~\ref{fig2}, also with a very small scatter. 
In that figure the factor $S_1$ is plotted -- for convenience -- as $C/S_1$, 
where $C$ is the integral of the rotational profile. The resulting curve is
very similar to the photometric light curve of AW~UMa.
The similarity of the shape is in fact expected (for an
invariable primary) as was discussed in Section~\ref{BFs} thanks to 
(1)~the built-in normalization of the BF integrals to
the integral of the template spectrum and (2)~the very weak 
variation of the apparent spectral type with orbital phase. 
Since the brightness and $V\sin i$ of the primary are apparently 
constant with phase, instead of the units resulting from the 
BF normalization (as in Figure~\ref{fig1}), 
we will use the central intensity ($S_1$) 
as a more convenient ``intensity unit'' (i.u.) in 
the subsequent plots and figures.

The results shown in Figure~\ref{fig2} were
obtained strictly spectroscopically, but they represent a very
reasonable light curve of AW~UMa which confirms the 
proper system of photometric phases used in this paper. The
figure also shows that most of the orbital phases were 
observed multiple times; the single-night coverage took place
only for two narrow phase intervals 0.98 -- 0.01 and 0.75 -- 0.80. 

During the phase interval $-0.07$ to $+0.07$,
the rotational-profile fits are distorted by the projection of 
the secondary component during its transit in front of of the primary star. 
This is reflected in the lower values of $S_1$ and by
the upward spike in the very centre of the primary eclipse in 
Figure~\ref{fig2}. The appearance of the
secondary component in the data is discussed further in 
Section~\ref{sec}.

\subsection{The rotation velocity}
\label{vsini}

\begin{figure}%[ht]
\begin{center}
\includegraphics[width=11.5cm]{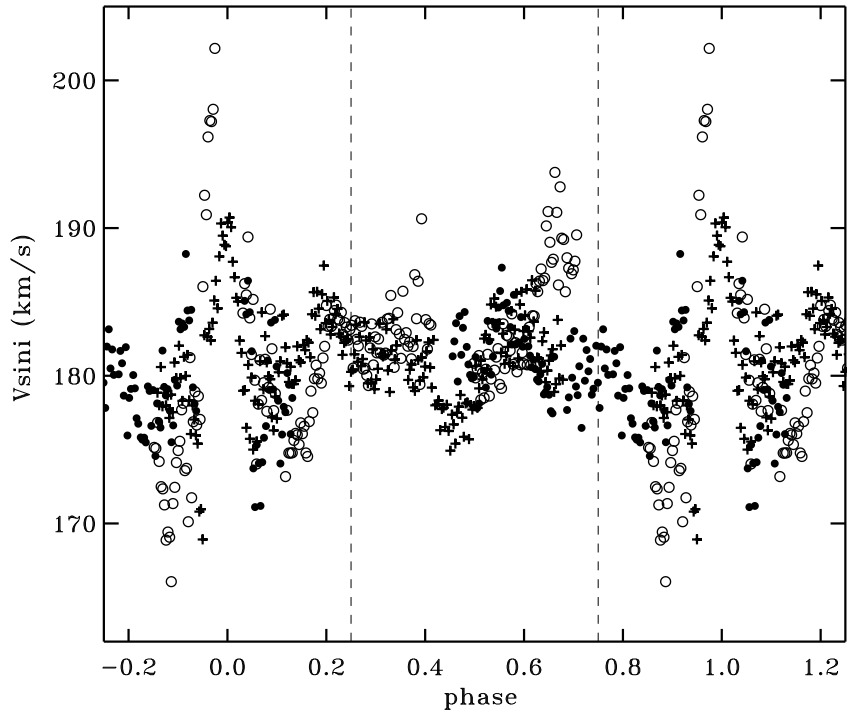}    
\caption{
\footnotesize 
The $V \sin i$ values determined by rotational-profile fits
to the upper 50\% of the primary component BF feature. 
The symbols for the three nights are the same as in 
Figures~\ref{fig2}. The median value of $V \sin i$ was determined 
within the phase range $0.25 < \phi < 0.75$.
}
\label{fig3}
\end{center}
\end{figure}

Values of the projected equatorial velocity
of the primary component, $V \sin i$, determined from the 
individual broadening functions are shown versus the orbital
phase in Figure~\ref{fig3}.
The median value which will be used below, 
$181.4 \pm 2.5$ km~s$^{-1}$, has been determined
for the phase interval $0.25 < \phi < +0.75$ (299 data points). 
The wide phase range around the primary eclipse, which was excluded,
resulted from indications that gas motions around the secondary 
component may distort velocities of the primary star 
(see Section~\ref{pri-orb}).
Some traces of systematic phase trends and nightly 
variations in $V \sin i$ indicate that gas motions may be present
at all phases and may influence the primary BF feature even within 
the upper 50\% of its height. 

\placefigure{fig3}      % fig.3

It has been long recognized that AW UMa is observed with 
an edge-on orbit, as implied by the long duration of the total eclipses. 
Thus, for solid-body rotation, the value of 
$V \sin i$ should be very close to the equatorial velocity of the primary.
Assuming that it rotates in synchronism with its orbital
motion, $V \sin i = 181.4$ km~s$^{-1}$ implies a radius of
$1.49\,R_\odot$, which is fully consistent with a F2V star.

It is possible that the actual $V \sin i$ is smaller than the
assumed median value, as indicated by a drop in its value to about
177 km~s$^{-1}$ at the very center of the secondary eclipse 
(phase 0.5, the small star behind), see Figure~\ref{fig3}. 
This would mean that the upper part of the 
BF profile is almost always broadened by gas streams.
However, variations in the individual determinations of
$V \sin i$ within the secondary eclipse 
between the individual nights force us to leave this issue as open
for now.

\subsection{The orbital motion of the primary}
\label{pri-orb}

\begin{figure}%[ht]
\begin{center}
\includegraphics[width=11.5cm]{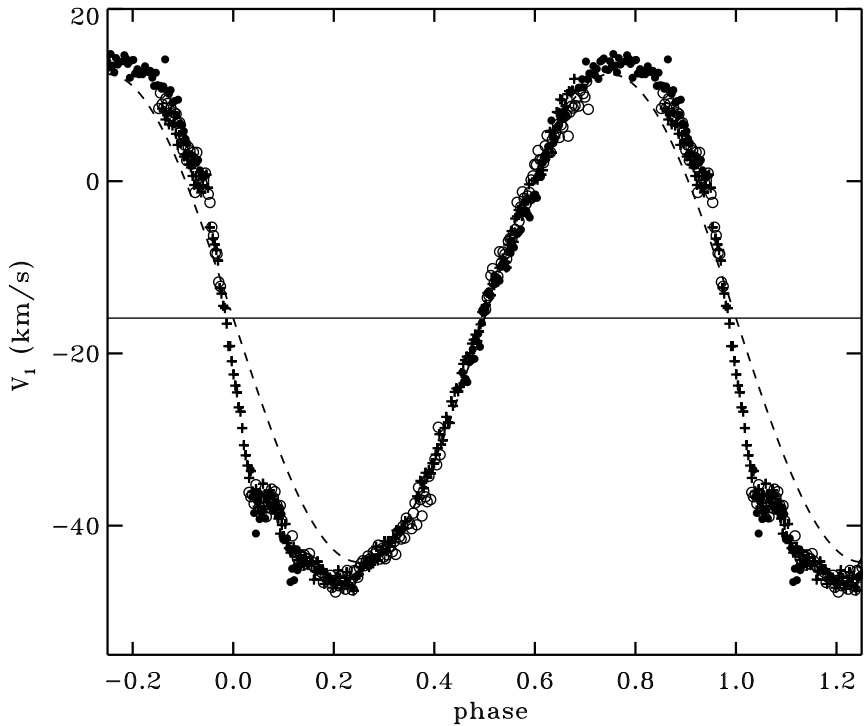}    
\caption{
\footnotesize 
Radial velocities of the primary component of AW~UMa 
determined by rotational-profile fits to the upper 50\% of the
primary feature. The symbols for individual nights are the same 
as in Fig.~\ref{fig2} and \ref{fig3}.
The broken line gives the sine-curve fitted in the phase range
$0.25 < \phi < 0.75$ with the parameters averaged from 
three sets based on the profile fits done at three different cutoffs
of the fitting rotational profile, as discussed in the text.
}
\label{fig4}
\end{center}
\end{figure}

The solution of the orbital motion of the primary component of AW~UMa 
was based on the 299 data points in the phase range 0.25 -- 0.75.
Within this range, the motion of the primary appears to be sinusoidal 
(see Figure~\ref{fig4}). 
Fits of the rotational profile to the primary component BF peaks may be
affected by complications due to the presence of the ``pedestal'' below the level
of about 0.4 of the BF maximum (see the next Section~\ref{pedest}) 
and by the surface inhomogeneities when the cut-off level for the fits
is set too high. It was determined that radial velocities 
measured with the low cut-offs within the range of 0.45 to 0.55 
of the maximum peak were least affected by both complications. 
Three sets of rotational profile fits were accordingly obtained 
for the upper 55\%, 50\% and 45\% of the observational BF's.
The results for all phases are tabulated in Table~\ref{tab_RVs}.
   
\placetable{tab_RVs}           % Table 3  ASCII with RVs, added in rev.
\placetable{tab_orb}           % Table 4  orbital elements, added in rev.

\begin{figure}%[ht]
\begin{center}
\includegraphics[width=12.5cm]{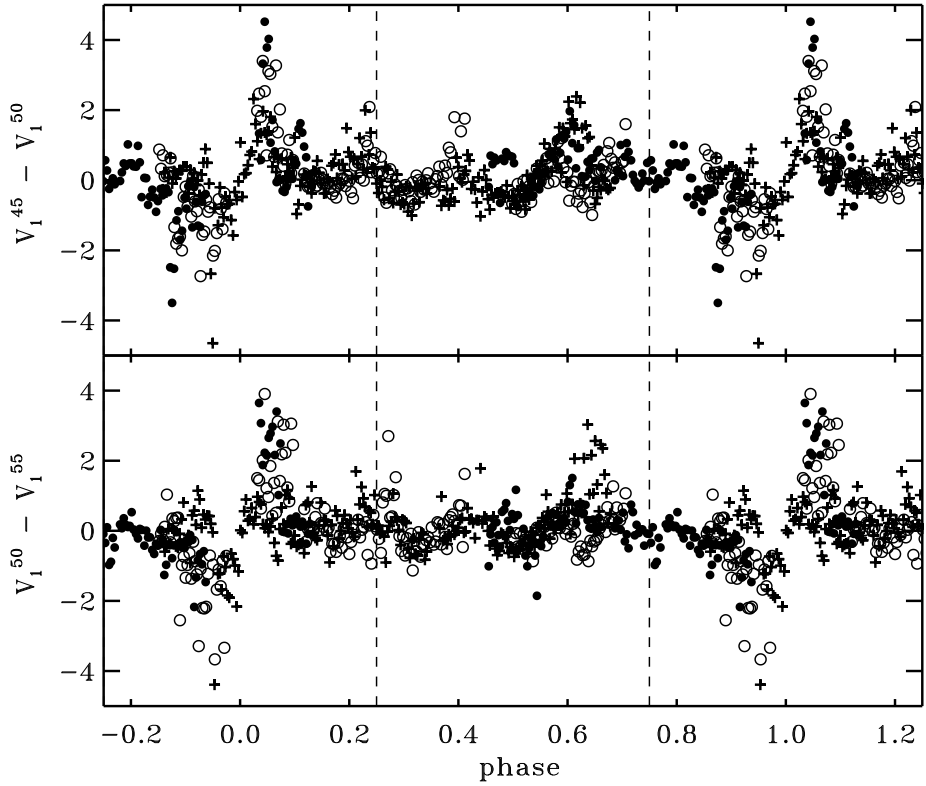}    
\caption{
\footnotesize 
Differences of the radial-velocity measurements of the AW~UMa
primary component, $V_1^{45} - V_1^{50}$ and 
$V_1^{50} - V_1^{55}$,
as measured by rotational-profile fits with a cut-off levels set at
0.45, 0.50 and 0.55 of the BF maximum. 
The symbols for the individual nights are the same 
as in the previous figures.
The vertical broken lines delineate the phase range $0.25 < \phi < 0.75$
used for three independent sine-curve fits. 
}
\label{fig5}
\end{center}
\end{figure}

The sine fits for the phase range 0.25 -- 0.75 for the three levels of
the cut-off are given in Table~\ref{tab_orb}. They are of high quality with 
the typical mean error of $\simeq 1.0$ km~s$^{-1}$ per point. Such a
measurement error, perhaps large by today's standards, 
is unprecedentedly small for a component of a W~UMa binary; 
when compared to the total width of the primary 
profile ($2 \times V \sin i$) it amounts to only 0.3\%. 
The individual radial velocity determinations for the three cut-off levels
are non-random and correlated (Figure~\ref{fig5}); obviously, these intrinsic
variations increase the above estimate of the fit error per point.
We see a systematic dependence in the derived values of $V_0$ and $K_1$ 
on the cut-off level; this dependence may possibly be modelled
once a correct model of the binary is developed.  
For the final orbital solution, we adopt the mean values of
$V_0$ and $K_1$, as listed in Table~\ref{tab_orb}, 
with the uncertainties estimated
from the scatter between the three sets of the parameters: 
$K_1 = 28.37 \pm 0.37$ km~s$^{-1}$, 
$V_0 = -15.90 \pm 0.12$ km~s$^{-1}$. 
The orbital amplitude $K_1$ is small when compared 
with the width of the primary profile: The star moves by only 7.8\% 
of its full width in radial velocities. 
The masses of the AW~UMa components resulting from the above
orbital elements are discussed in Section~\ref{orbit}. 

\placefigure{fig4}      
\placefigure{fig5}      %  new figure

The RV data deviate from the sine curve over the wide range 
of orbital phases close to the primary eclipse 
(at least within $-0.8 < \phi < +0.2$), which implies 
that the radial velocity curve of the primary component is
perturbed by the projection of the secondary component. 
The perturbations show some similarity to the
``Rositter - McLaughlin effect'' which 
-- in terms of the RV amplitude -- was noted by \citet{Wilson2008}, 
on the basis of much less detailed observations, to be unexpectedly 
small or perhaps entirely absent. 
Our observations confirm that the eclipse disturbance is
indeed much smaller than predicted by Lucy's contact model.
% used by \citet{Wilson2008} and several previous investigators. 
However, it extends in the orbital plane far from the centre
of the secondary component. In fact, judging by the extent of the
perturbation, the secondary appears to be 
even larger than the primary component. 
This may be taken as one of several
indications that the secondary is not a star filling
its Roche lobe but rather a flattened structure. 
The radial-velocity disturbances within the primary eclipse
($-0.8 < \phi < +0.2$) are asymmetric relative to the line 
joining the stars;  a distinct isolated feature 
around the phase +0.05 was observed on all three nights,
see Figure~\ref{fig4}.

\begin{figure}%[ht]
\begin{center}
\includegraphics[width=11.5cm]{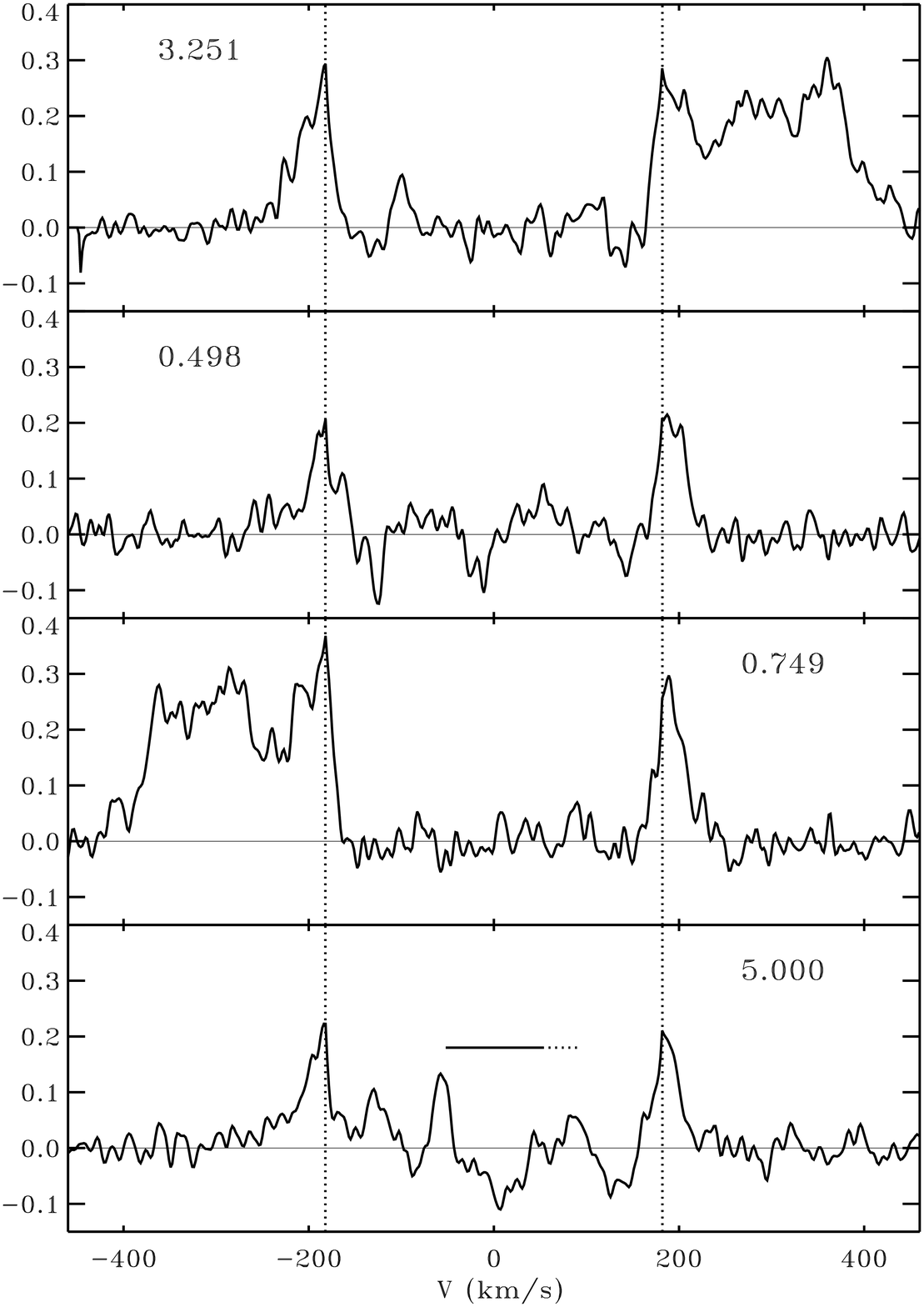}    
\caption{
\footnotesize 
Deviations from the fits of the rotational profile to the 
primary component are shown here for the same phases as in
Figure~\ref{fig1}. The ``pedestal'' is the additional light
at and beyond the limit defined by $V \sin i = 181.4$ km~s$^{-1}$
(marked by the vertical dotted lines).
The velocity system is shifted to the centre of the primary component. 
In this and in the subsequent figures the vertical axis is in 
intensity units (i.u.) of the central brightness of the primary component. 
The range, where the secondary component appears to be projected onto the
primary star (see Section~\ref{sec}), is shown by a horizontal bar 
in the last panel.
}
\label{fig6}
\end{center}
\end{figure}

\placefigure{fig6}

\subsection{The pedestal}
\label{pedest}

\begin{figure}%[ht]
\begin{center}
\includegraphics[width=11.5cm]{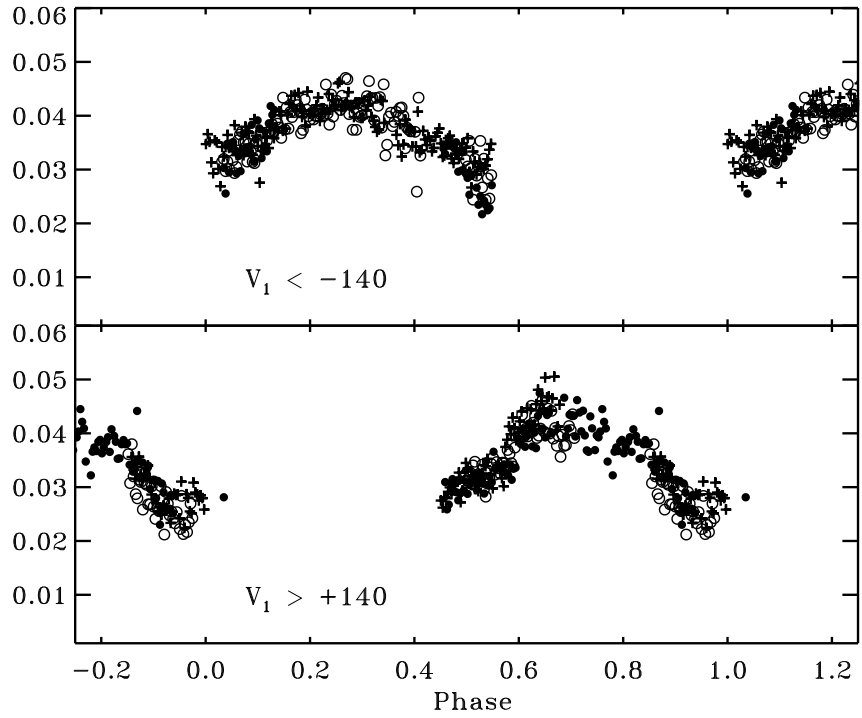}    
\caption{
\footnotesize 
The flux of the pedestal integrated between
velocities $-250 < V_1 <-140$ km~s$^{-1}$ (the upper panel)
and $+140 < V_1 < +250$ km~s$^{-1}$ (the lower panel), expressed
relative to the integrated flux within the rotational profile
of the primary component.
The symbols used for the three nights are the same as in the
previous figures.
}
\label{fig7}
\end{center}
\end{figure}

An additional RV component is observed at radial velocities 
of about 150 -- 250 km~s$^{-1}$ relative to the center of 
the primary component. It is called 
the ``pedestal'' because it presents, in Figure~\ref{fig1},
as an extended base of the central rotational profile. 
It is best seen when the same rotational profile (assumed
$V \sin i = 181.4$ km~s$^{-1}$ and scaled by $S_1$, 
see Section~\ref{pri-fits}), 
is subtracted from individual BF's, as in  Figure~\ref{fig6}. 
The pedestal was discovered by \citet{PR2008}. No 
interpretation was offered there except for noticing 
that it must be produced by an optically 
thick material (to produce the same absorption spectrum as the star) 
moving around the primary at high rotational/orbital velocities. 
While it is observed at both negative and positive 
velocities relative to the center of the primary, it is better defined 
for velocities opposite to those of the secondary component.
The phase dependence of the integrated intensity of the pedestal, 
after removal of portions of the orbit
which can be affected by the secondary star, is shown in 
Figure~\ref{fig7}. The pedestal intensities, when expressed 
in units of the integrated rotational profile,
vary at the level of 2.5\% to 4\% with the phase,  
following the double cosine curve, as expected for an elliptical 
distortion. The width of the pedestal does not seem to be phase 
dependent and stays within about 50 km~s$^{-1}$ beyond the
rotational profile, so that the variations are mostly 
due to the changing brightness within it. 

At this point one cannot say if the pedestal originates 
{\it on the primary component\/} as a belt of rapidly-moving 
gas or a disk in the orbital plane, but {\it outside
the star\/}. The moderate intensity of the pedestal suggests that this
distinction is almost immaterial and that the gas must be rather
strongly confined to the equatorial plane.

\placefigure{fig7}        

\placefigure{fig8}      
\placefigure{fig9}      
\placefigure{fig10}

\section{Surface inhomogeneities on the primary component}
\label{inhom}

\subsection{Types of features}
\label{types}

The two-dimensional images of the velocity field over the disk of
the primary component of AW~UMa are presented in 
Figures~\ref{fig8} -- \ref{fig10}. The figures show 
surface features which are not expected for a simple rotating star. 
The figures have been obtained by (1)~shifting the velocities to the
center of the primary component using the value of $K_1$
(Section~\ref{pri-orb}), (2)~subtracting the best-fitting 
rotational profiles from the BF's and (3)~re-sampling the 
individual deviations into a uniform phase
grid at the step of 0.0033 in phase (125 sec in time).

In addition to the pedestal discussed in Section~\ref{pedest}, 
which presents in the figures as two vertical ridges along the edges of
the rotational profile, one sees two main types of 
feature on the surface. We call them ``ripples'' and ``spots''. 
Both of them appear to follow the rotational motion of the 
primary, but several details allow to distinguish them in 2D figures,
so we discuss them separately. 

\subsection{Ripples} 
\label{ripp}

\begin{figure}%[ht]
\begin{center}
\includegraphics[width=12.5cm]{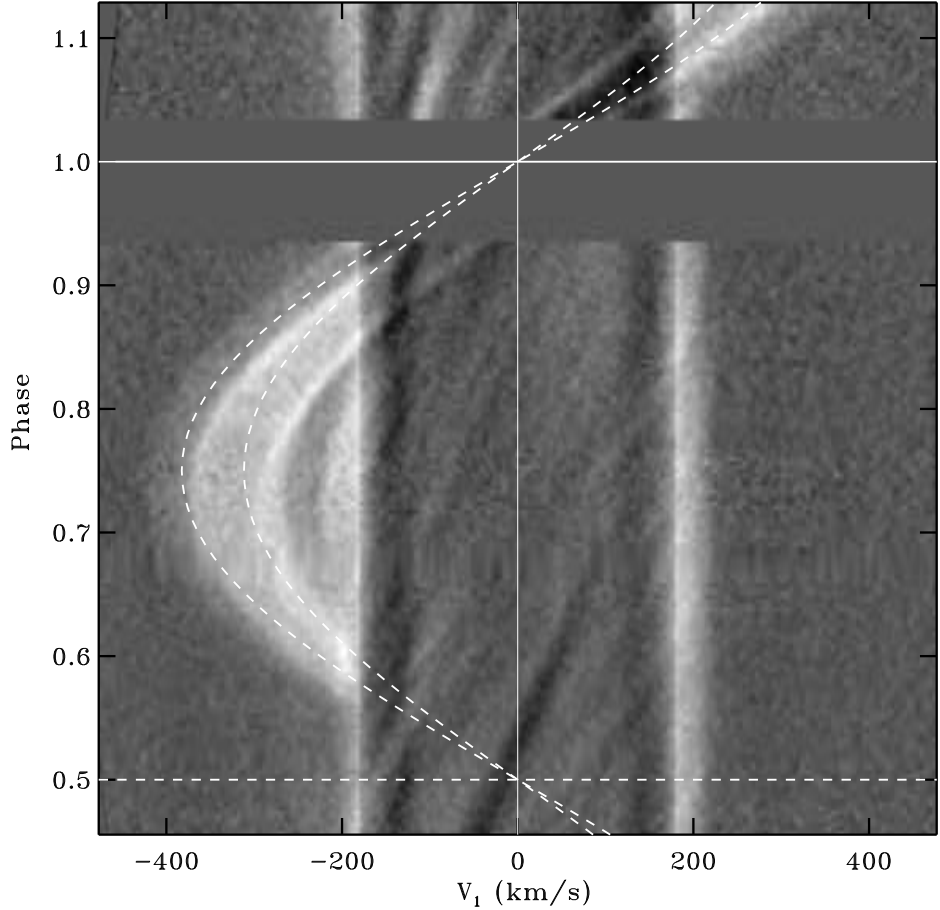}    
\caption{
\footnotesize 
The two-dimensional representation of variations
in the Broadening Functions of AW~UMa versus the orbital phase
for the first night of the CFHT run.
The horizontal axis is the velocity relative to the center of
the primary component while the vertical axis is the orbital phase
counted uniformly through the observing run. The BF's for the
primary component have the same (scaled in amplitude) rotational profile 
$V \sin i = 181.4$ km~s$^{-1}$ subtracted to
improve the visibility of the surface inhomogeneities. The dark, strongly
inclined feature at the bottom of the figure 
($\phi \simeq 0.5$, $V_1 \simeq 0$ km~s$^{-1}$) is a spot. The
ripples form a tenuous grid of slightly less steeply inclined, 
weaker lines.
The continuous broken lines represent the expected motion of
the secondary component for two mass ratios, $q = 0.10$ and 0.08;
the amplitude is larger for the smaller $q$. The grey area close
to phase 1.0 in the upper part of the figure corresponds 
to phases which were not observed.
}
\label{fig8}
\end{center}
\end{figure}

\begin{figure}%[ht]
\begin{center}
\includegraphics[width=12.5cm]{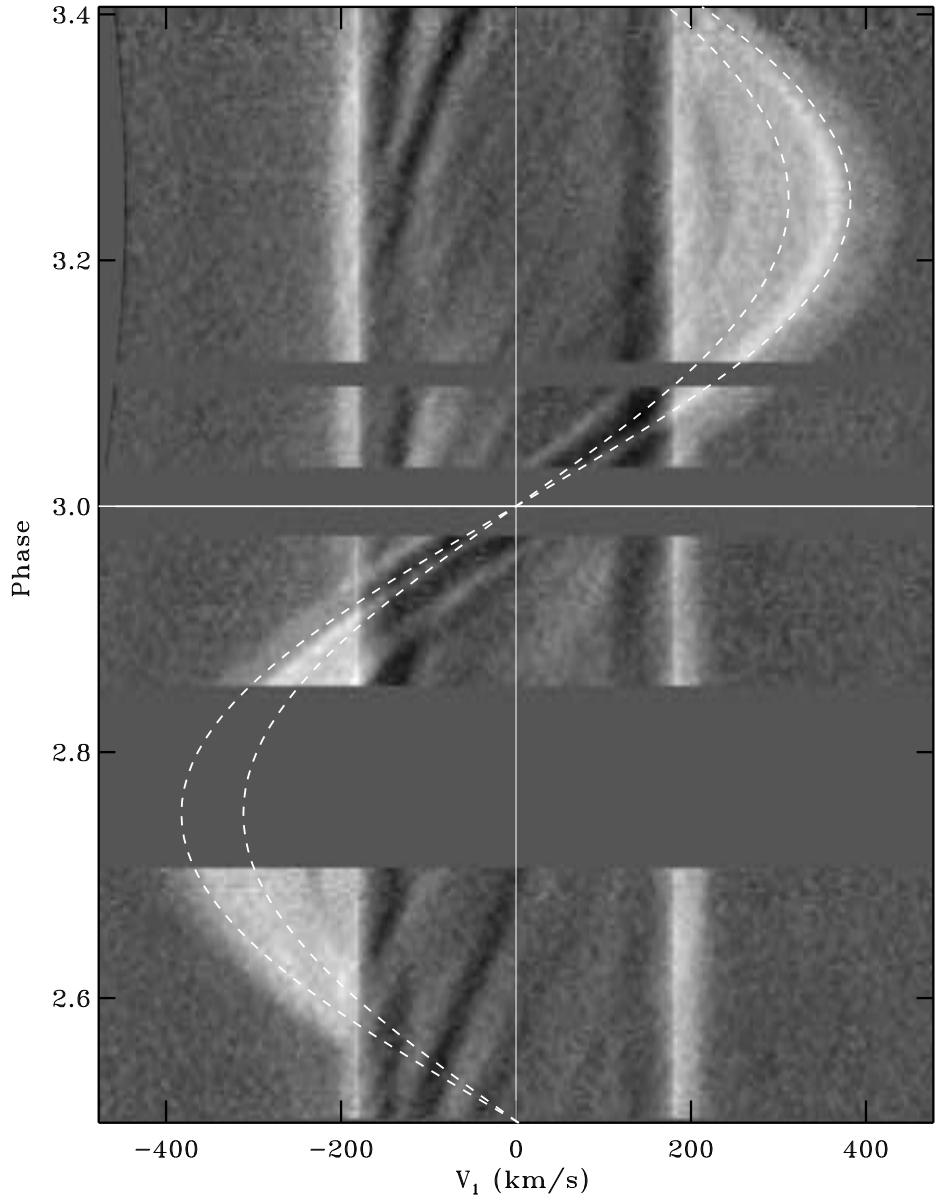}    
\caption{
\footnotesize 
The same as for Figure~\ref{fig8}
for the second night of the CFHT run.
}
\label{fig9}
\end{center}
\end{figure}

In the image of deviations from the rotational profiles in
Figures~\ref{fig8} -- \ref{fig10}, the ripples form 
a grid of tenuous, inclined lines crossing the whole 
range of velocities which are related to the primary component.
They are narrow with the width approaching the RV resolution 
of our data. The ripples can be distinguished from 
the spots by their velocity dependence: The spots 
tend to curve towards the stellar disk edges (somewhat 
similarly to the $\arcsin$ function) and do not extend
beyond $\pm V \sin i$; in contrast,
some of the ripples appear to continue into the pedestal, i.e.\ 
outside the stellar disk. Thus, the gas responsible 
for the formation of the pedestal and the ripples
may have the same origin. The ripples are surprisingly
straight which may indicate that the physical structures
causing them are located above the stellar surface.
%When analysing the ripples, we did not apply any corrections
%for the projection shortening towards stellar edges as this
%would distort what is directly visible. 

Four rectangular sections of 
Figures~\ref{fig8} -- \ref{fig10},
covering phases 0.60 -- 0.87, 3.12 -- 3.30,
5.09 -- 5.34 and 5.36 -- 5.51 and at the center of the disk,
within $\pm 130$ km~s$^{-1}$ in velocities were analyzed 
for the frequency content in the deviations from the rotational 
profile. They were selected so as to avoid obvious spots or traces 
of the secondary transiting the primary. For each section,
the FFT analysis was performed on all available observational BF's,
ranging in number between 42 and 72 per segment.  The median
amplitudes for each segment are plotted in the upper part of 
Figure~\ref{fig11} versus the spatial (the RV space) frequency 
expressed in the language of non-radial pulsations as
$l$-degree modes. The amplitudes appear to 
rise toward low frequencies (i.e.\ larger RV scales), but
cannot be determined for $l < 30$ because of the finite extent 
in the radial velocity range and the presence of spots. At the
lowest accessible spatial frequencies the amplitudes reach 
$\simeq 0.008$ i.u. Thus, the ripples are faint, and their detection 
at levels even below 0.001 i.u. has been only possible 
thanks to the high quality of the CFHT observations. 
The frequency content of the ripple amplitudes
partly reflects the wispy structure of the thin ripples as the 
individual strands sometimes show larger contrast with 
intensities $>0.01$ i.u. 

The ripples may possibly be both bright and dark, but this
obviously depends on the way how the rotational profile
is fit and subtracted. If some are bright, as it seems for a few
extending into the pedestal, this property would
distinguish them from the dark spots. 
Taking into account the properties of the BF's
(Section~\ref{BFs}), the bright strands may in fact have a
lower temperature than the surroundings as then BF features
become stronger in intensity (provided the matter remains 
optically thick). 

\placefigure{fig11}              

The best defined ripples with amplitudes larger than about 
0.005 -- 0.007 i.u.\  (depending on limits set by the local confusion)
were analyzed in 2D images (Figures~\ref{fig8} -- \ref{fig10}), 
for their duration (expressed in phase $\phi$) and the
slope $dV/d\phi$ at the center of the primary profile,
i.e.\ at the meridian crossing point. 
While we do not see any regularities in the durations,
the slopes (counted per one rotation synchronized with the
orbital period) give some information on typical 
drift velocities across the stellar disk. 
The lower panel of Figure~\ref{fig11} gives the histogram 
of the drift velocities for the 35 measured, stronger ripples. 
While the accuracy is low,
typically $\pm 25$ km~s$^{-1}$ and there may exist a genuine 
spread in the ripple slopes, the median 
$dV/d\phi/2\pi = 170 \pm 21$ km~s$^{-1}$ is consistent
with the projected equatorial velocity, $V \sin i = 181.4$ km~s$^{-1}$.
Thus, the ripples appear to participate in the rotation of the
primary component. Their straight shape and extensions into
the pedestal are then hard to explain as surface features;
they may very well be structures above the surface, but sharing 
the rotation of the star.

\subsection{Spots}
\label{spots}

Spots are inhomogeneity features on the primary component 
which are dark in the BF's and appear within $\pm V sini$, 
but do not extend into the pedestal. 
Thus, they seem to be confined to the range of the
surface radial velocities.  

The spots are always best visible at the negative-velocity side of the
BF's where they can be as deep as 0.1 i.u. 
As can be seen in Figures~\ref{fig8} -- \ref{fig10},
the spots emerge at negative velocities slowly, curving 
upward as expected for surface features and then accelerating in
radial velocities (turning more horizontal) towards the 
sub-observer meridian. Their apparent drift 
at the meridian crossing is unexpectedly slow: In the 2D images they
are more vertical than the synchronous rotation would
imply; if they shared the surface rotation as given by $V \sin i$,
then their horizontal drifts would be much faster across the profile. 
The mean drift velocity at the meridian is relatively uncertain, 
$dV/d\phi/2\pi = 120 \pm 30$ km~s$^{-1}$, because most spots
do not cross the central meridian to reach positive 
velocities; usually they disappear while still 
approaching the observer.

\placefigure{fig12}            

If interpreted as rotational velocities, the slow drift velocities
at the meridian of $\simeq 120$ km~s$^{-1}$ 
would imply a rotation period about 3/2 longer than 
the one resulting from $V \sin i$ and the synchronous rotation rate. 
Such a surprisingly slow rate finds some support in an attempt to phase
the reoccurrence of the spots, as shown in Figure~\ref{fig12}.
In this figure the spot drifts are shown for the periods which
appear to give the best clustering of the individual lines, of 1.00, 1.46
and 1.52 times $P_{\rm orb}$. While an exact alignment of the tracks is
hard to achieve for any period tested, the scatter visibly 
diminishes for the slow rotation rates of around $3/2 \times P_{\rm orb}$.
Such a slow rotation rate is unlikely, given the
excellent fits of the rotational profile to the upper parts of
the primary star BF profiles. Thus, we may suspect that the spots
are due to surface structures moving {\it relative\/} to 
the stellar surface with their own, slow rotation rate. 
The Doppler mapping of such features would be very
difficult to achieve for the short rotation period of AW~UMa
and if the differential rotation rate is indeed as large 
as $\simeq 50$\%. 

The lack of an obvious Stokes $V$ signal (Section~\ref{intro})
suggests that spots are non-magnetic structures, but it is 
difficult to determine what is the nature of the 
spots and why do they appear dark. 
They may be genuinely dark, but it is hard to imagine photospheric
spots which would lower the spectral continuum locally but
have not already been detected in photometry (although a comparison of 
the light curves by \citet{Prib1999} does show significant seasonal 
changes in AW~UMa light curves). 
More likely the spots are visible only spectroscopically
and possibly only through the use of the spectral deconvolution.
This would happen if the spectral lines are locally 
shallower than expected, as then depressions in the BF's would appear. 
The lines could become more shallow due to filling-in by emission
or to locally higher gas temperatures.
This matter remains unresolved at this point. 

Irrespectively of the period assumed for the spot movement rate,
synchronized with $P_{\rm orb}$ or equal to $3/2 P_{\rm orb}$, as in
Figure~\ref{fig12}, 
the spots tend to appear mostly at the sub-secondary and the opposite
``ends'' on the primary component, along the line
joining the stars. This may be due to 
instabilities and/or condensations in flows predicted by \citet{Oka2002}
for the mass-loosing component of a semi-detached binary.
According to these model computations, the
flow should take place from the high-pressure polar regions 
to the low-pressure equatorial regions with the latter 
situated along the line joining the mass centers. The 
high-pressure counter-rotation in the polar regions in the model 
of \citet{Oka2002} may have links to the puzzling visibility 
of the spots only at the negative-velocity side: If the spots are
extended but thin, they may be curved in such a way as to 
be visible at similar velocities at the negative side, but distributed
in velocities (hence harder to detect) 
while on the positive side of the profile.

\section{The secondary component}
\label{sec}

\subsection{The velocity field around the secondary}
\label{sec-vel}

AW~UMa shows a systematic period change 
$dP/dt = -5.3 \times 10^{-10}$ or $-1.9 \times 10^{-7}$ day/year
\citep{Rci2013b} which is typical for about a quarter of  
W~UMa type binaries for which such changes have been detected
\citep{Kubiak2006}; most of them show smaller, randomly distributed
period changes. When interpreted
as resulting from the conservative mass transfer from the more massive
to the less massive component, the mass exchange rate (estimated using
Eq.~4.56 in \citet{hilditch}) is: 
$dM/dt = (1/3) \, (M_1 M_2)/(M_1-M_2)\, (dP/dt)/P
               \simeq -2.1 \times 10^{-8} M_\odot/{\rm yr}$.
Note that this is the {\it net mass transfer rate\/}: The actual
amount of the mass flowing from the primary to the secondary may be 
larger, as part of the flow may return to circle around the
primary component and feed the pedestal.
The mass-transfer process leads to accretion phenomena 
clearly visible spectroscopically on and around the secondary component.  
According to \citet{PR2008}, the secondary may actually be
an accretion disk or a star submerged in a disk-like structure. 

Figure~\ref{fig13} shows profiles obtained by averaging nine
individual BF's observed within $\pm 0.015$ in phase around each of the
three fully observed orbital quadratures. 
The secondary shows a double-peaked shape 
which is characteristic for an accretion region. The
shape changes from night to night in intensity and in 
spacing between the peaks. For the three observed quadratures, the peak
separation (measurable to about $\pm 3$ km~s$^{-1}$)
varied distinctly: 80, 98 and 65 km~s$^{-1}$ with 
the mean half-separation of $40.5 \pm 4.8$ km~s$^{-1}$.
The outer wings, beyond the peaks, outside of the 
inner $\pm (60 - 70)$ km~s$^{-1}$, were the same for the two quadratures 
when the secondary was moving away, but they were quite different
for the single quadrature when the secondary was approaching us;
this was due to the influence of the pedestal which was apparently 
different when seen from both sides of the binary. 
The mean velocities (relative to the primary mass center),
determined as averages from the peak positions
at the three quadratures, are surprisingly similar: 
$-320.6$, $+310.8$, $+311.2$ km~s$^{-1}$. This gives the mean 
$K_1 + K_2 = 314.2 \pm 3.2$ km~s$^{-1}$. This number, taken  
with $K_1 = 28.37 \pm 0.37$ km~s$^{-1}$, leads 
to the mass ratio $q_{\rm sp} = M_2/M_1 = 0.0991 \pm 0.0014$. We will use
the values of $K_1$ and $q$ for determination of the component masses in
Section~\ref{binary}. Here we note that the error of
$K_1 + K_2$ as given above may be fortuitously small as it is based 
on only three determinations; 
we will use $\pm 0.003$ as the error of $q_{\rm sp}$.

A major difference in the interpretation results from two possible
assumptions on what we actually see as the secondary component, 
an accretion disk or manifestations of accretion processes 
on the surface of a star. For a stable accretion disk, the 
distinct peaks in the RV profile are produced by the slow-moving 
material at the outer rim of the disk, while the extended wings outside 
of the peaks are formed by the large-velocity matter in the
innermost parts of the disk \citep{smak1981,smak1993}. To be
visible in the BF's, such a disk would have to consist of optically 
thick matter to produce a stellar spectrum with the same
absorption lines as for the stellar surface. Obviously, the geometry
of an optically thick disk would require an axis
inclination of $i \ne 90$ degrees. By contrast, for
a star undergoing accretion, the observed effect 
would entirely depend on processes within a boundary layer 
between the stellar surface and the incoming material. The
latter possibility, of an interaction between the star and 
some sort of quasi-disk, seems to be more likely. In addition, 
the accretion-disk interpretation encounters the main difficulty 
in the expected size of the disk: 
For a stable Keplerian disk with the central object of such a small
mass as the secondary of AW~UMa, the observed separation between
the velocity peaks implies disk dimensions of the order of $15\, R_\odot$
(this results from a simple scaling relative to the Earth velocity:
$V_d \propto 30\, (M/M_\odot)^{1/2} (215 R_\odot/R)^{1/2}$ km~s$^{-1}$).
There is no space to accommodate such a large disk. However,
we do see -- through the {\it quasi-}Rossiter-McLaughlin effect 
(Section~\ref{pri-orb}) -- that gas extends as far from the 
secondary component in the orbital plane as perhaps to a distance
$\simeq 1\,R_\odot$. Thus, most likely, the observed ring-like structure 
around the secondary component is neither a fully developed 
Keplerian disk nor a boundary layer on top of the star. 
It may be an interaction region where the 
returning matter, which had missed the secondary and
already made a full revolution around
the primary encounters the one orbiting the secondary component.
The losses due to the eventual accretion are replenished by the new
matter coming from the relatively nearby $L_1$ Lagrangian point of
the primary component. The amount of the dispersed matter in the
orbital plane may be appreciably larger than the rather moderate
net mass transfer rate as indicated by the period change. 

\placefigure{fig13}        
\placefigure{fig14}

The complicated accretion flow around the secondary component 
is visible when velocities are shifted to the expected 
mass centre for the assumed orbital parameters of the binary. In 
Figure~\ref{fig14}, the 2D images show the secondary component
with velocities shifted to its expected mass center for the assumed
$K_1 = 28.37$ km~s$^{-1}$ and for two values of the mass ratio, 
$q_{\rm sp}=0.08$ and 0.10. 
The motion of the secondary star is very sensitive to $q_{\rm sp}$, 
so that images such as Figure~\ref{fig14} provide a direct check 
on the assumed value of $q_{\rm sp}$. The double-peaked structure,
which is well defined during the three orbital quadratures 
(Figure~\ref{fig13}), does not seem to continue as
a simple, vertical band in this picture for any of the assumed 
values of $q_{\rm sp}$, but shows complex changes in brightness and position.
The mass ratio $q_{\rm sp} \simeq 0.10$ seems to produce a more stable 
profile, but there still exist complex motions within the profile
which are not in strict anti-phase relation to the primary component. 
Unfortunately, even the almost-continuous monitoring on three nights 
did not provide sufficient coverage to establish any 
regularity in those strands and wobbling motions.
Thus estimated mass ratio, $q_{\rm sp}=0.10$, is basically identical 
to that determined from the 
mean positions of the double peak at the orbital quadratures,
$q_{\rm sp} = M_2/M_1 = 0.099 \pm 0.003$, where the assumed
uncertainty is twice as large as the one formally determined from
the mean velocity of the double-peaked structure. 

The secondary appears to look much more stable when all the data from
the three nights are binned in phase and velocity, 
as is shown in Figure~\ref{fig15}. 
The individual strands within the accretion structure
average out in such an image. The secondary 
appears to have a somewhat trapezoidal shape bordered by two ridges;
the structure follows the expected sine-curve motion of the
secondary component for $q_{\rm sp} \simeq 0.10$.
In this average, binned image of the whole AW~UMa system,
the parts which undergo most variability are located
in the region of the two peaks in the profile of the secondary component
(Figure~\ref{fig16}); most of the binary is stable and repeatable. 
We note that the primary component, except for its surface ripples and
spots, is practically invariable anywhere in its velocity field. The
pedestal feature is also surprisingly constant with respect to time.

\placefigure{fig15}          
\placefigure{fig16}

\subsection{Dimensions of the secondary}
\label{sec-size}

Eclipses have a potential of shedding light on the physical 
state of AW~UMa secondary component.
However, in attempting an interpretation, one must remember
that the RV data do not tell us anything about 
geometry of the component stars, but only about
the velocities involved. Thus, the shapes in the 2D images 
%presented above 
require assumptions on how velocities relate to spatial positions. 
For the secondary component, any interpretation carries an important
uncertainty related to the ``inside-out'' visibility of accretion 
processes where the largest velocities are expected closest 
to the star, rather than furthest from it, as expected 
for rapidly rotating stars. 

\begin{figure}%[ht]
\begin{center}
\includegraphics[width=12.5cm]{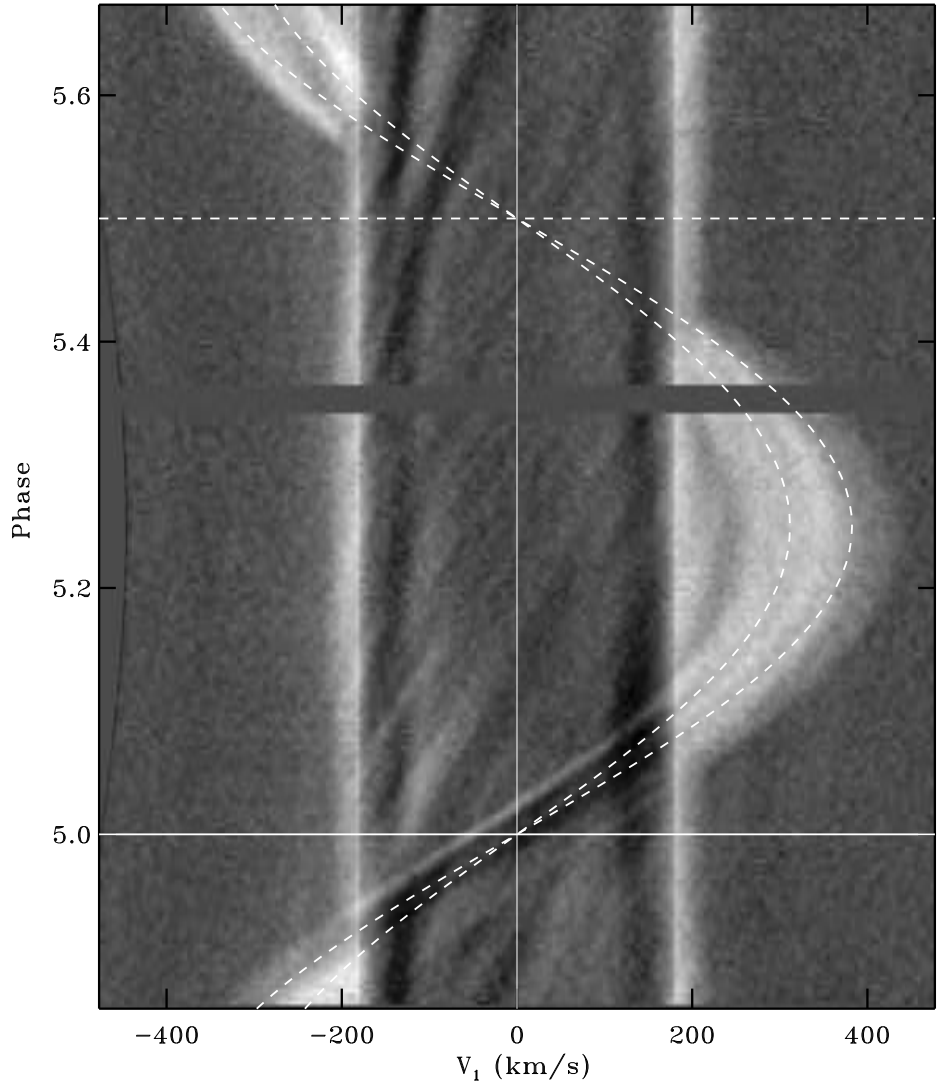}    
\caption{
\footnotesize 
The same as for Figure~\ref{fig8}
for the third night of the CFHT run.
}
\label{fig10}
\end{center}
\end{figure}

\begin{figure}%[ht]
\begin{center}
\includegraphics[width=9.5cm]{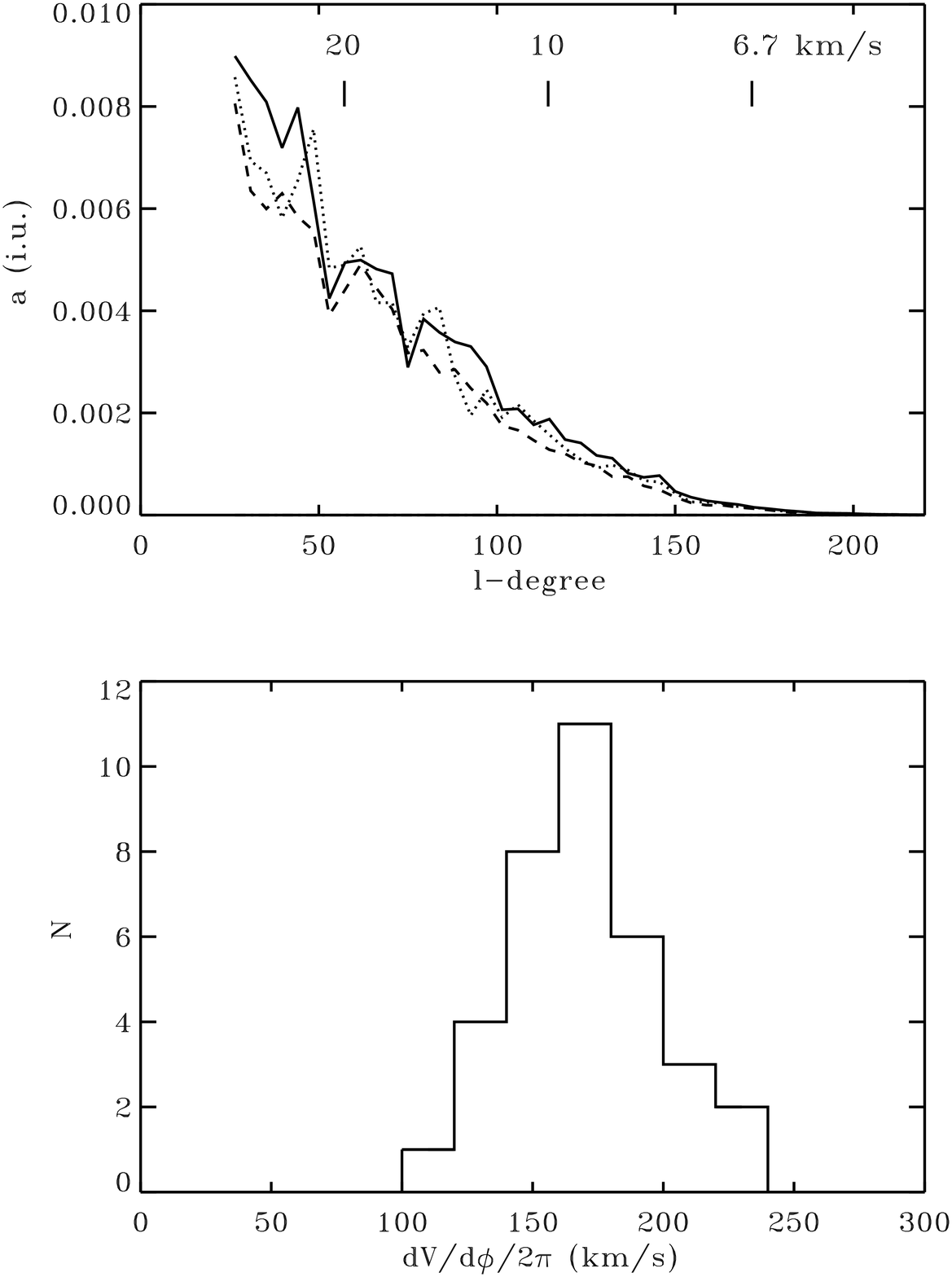}    
\caption{
\footnotesize 
The upper panel shows the amplitudes of spatial frequencies 
(in radial velocity space) for the surface ripples within 
four phase sections, as described in the text. The horizontal
axis is in the angular degrees, $l$, as typically used for description
of non-radial pulsations.
The intensity unit (i.u.) for the amplitudes 
is the brightness of the primary star at the  centre of its BF profile. 
The size of the corresponding features in radial velocities
is given along the upper horizontal axis. Note that the resolution
of the BF's was set at 8.5 km~s$^{-1}$ so that fluctuations 
with $l > 135$ represent noise. The lower panel gives
the histogram of the drift velocities at the central meridian for the
best defined ripples, typically seen at the contrast of 
at least 0.005 -- 0.007 i.u.
}
\label{fig11}
\end{center}
\end{figure}

\begin{figure}%[ht]
\begin{center}
\includegraphics[width=11.5cm]{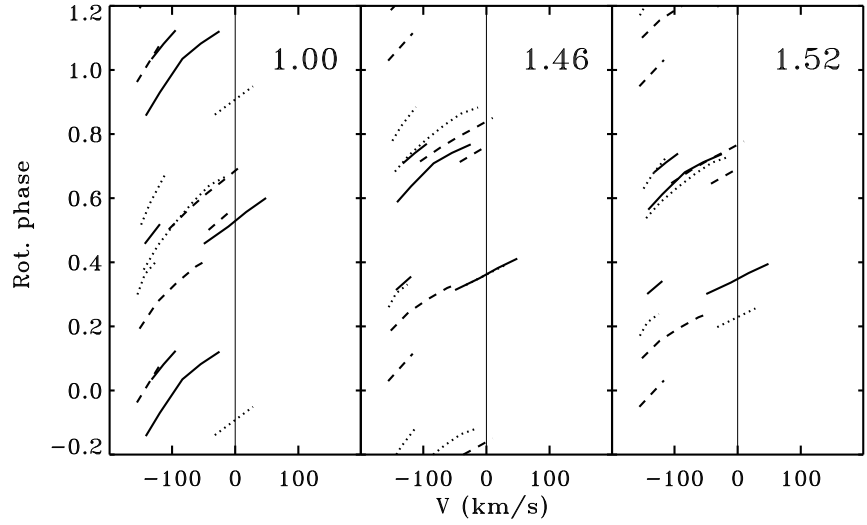}    
\caption{
\footnotesize 
The spot drift diagrams in the velocity vs.\ the rotational phase 
for the assumed three rotation periods of the spots: $P_{\rm orb}$,
$1.46 \times P_{\rm orb}$ and $1.52 \times P_{\rm orb}$. 
The lines give the spot positions for the three nights:
continuous -- night \#1; dashed -- night \#2; dotted --
night \#3. The spots usually disappeared before the
meridian transit seldom reaching positive velocities.
}
\label{fig12}
\end{center}
\end{figure}

\begin{figure}%[ht]
\begin{center}
\includegraphics[width=12.5cm]{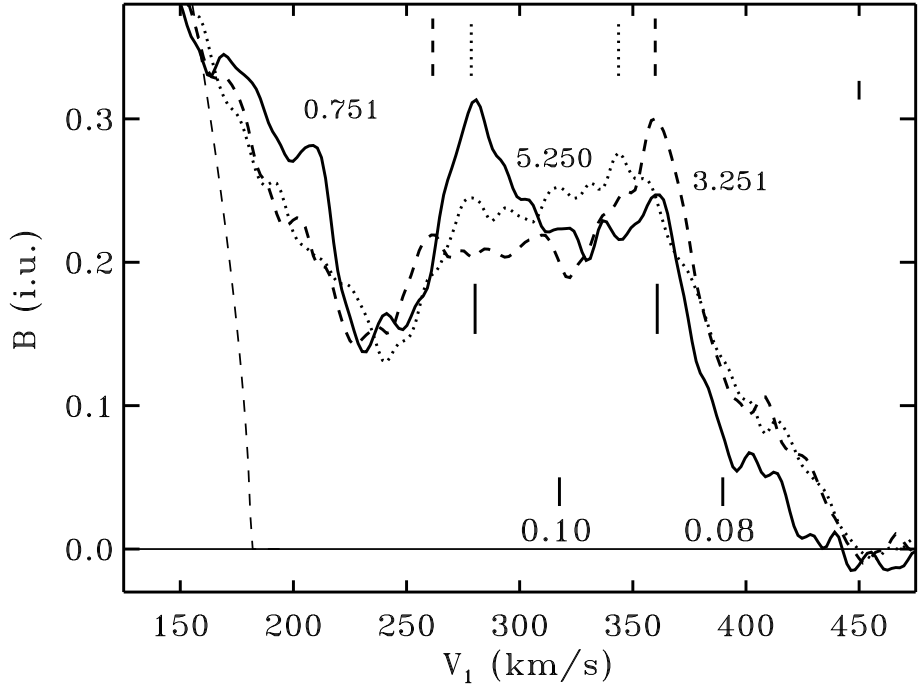}    
\caption{
\footnotesize 
The secondary component, as visible in radial velocities 
within $\pm 0.015$ of the three observed orbital quadratures.
The velocities are relative to the primary component; 
the steep, broken, thin line at the left of the figure
gives the rotational profile of the primary.  
Of the three quadratures, two were observed
with the secondary receding from the observer (the nights
\#2 and \#3, phases 3.251 and 5.250) and one quadrature
was observed with the secondary approaching the observer 
(the night \#1, phase 0.751; for this night 
the velocity scale was inverted).  The different lines are used:
continuous -- night \#1; dashed -- night \#2; dotted --
night \#3. The pairs of corresponding vertical bars
mark positions of the accretion profile peaks on those nights. 
The error bars resulting form averaging of
9 individual BF's for each case are not shown for
clarity of the figure; the median value of the error, 0.0065 i.u., 
is shown as a short vertical bar in the upper right corner.
The markers 0.08 and 0.10 in the lower part of
the figure give the predicted positions of the secondary 
centroid (i.e. $K_1+K_2$) for the measured amplitude 
$K_1$ and the two values of the mass ratio $q$.
}
\label{fig13}
\end{center}
\end{figure}

\begin{figure}%[ht]
\begin{center}
\includegraphics[width=11.5cm]{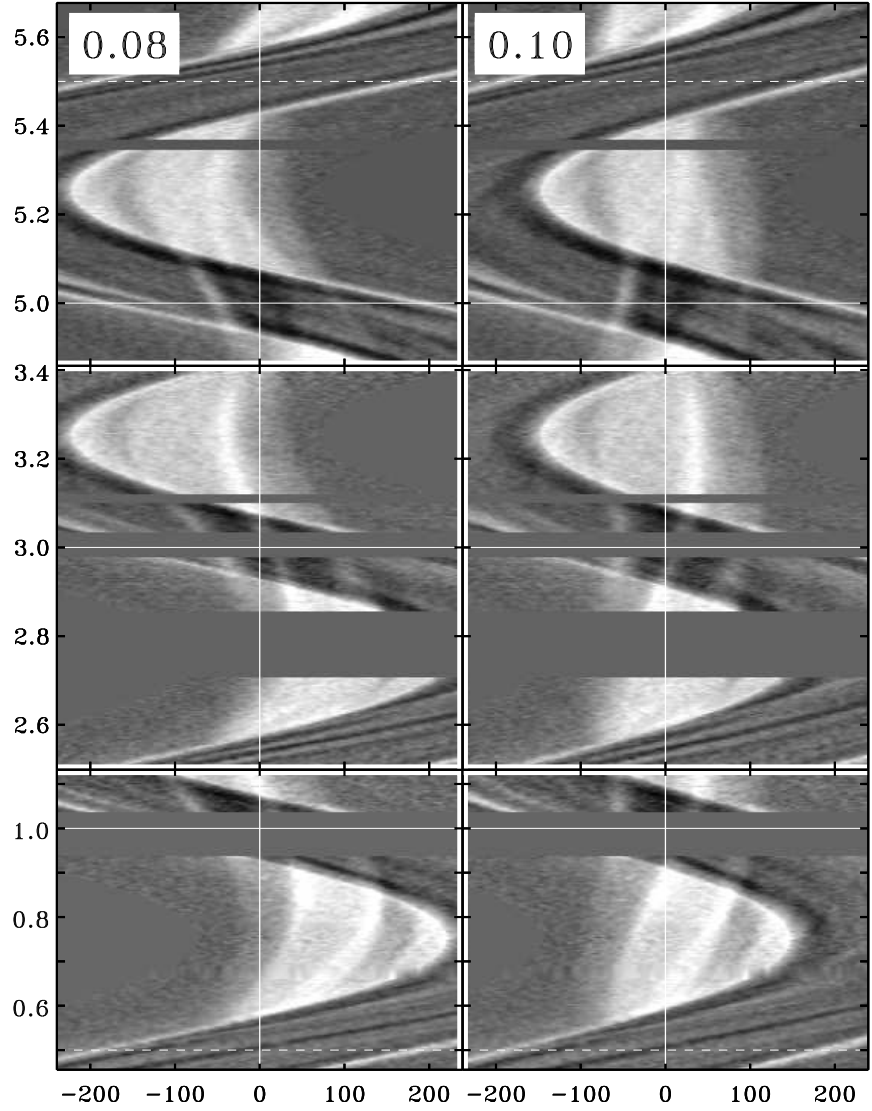}    
\caption{
\footnotesize 
The two-dimensional images of the secondary component of AW~UMa
in velocity space, similar to 
Figures~\ref{fig8} -- \ref{fig10},
but with the velocity system shifted to the expected motion
of the secondary component centroid for two values of the mass
ratio, 0.08 and 0.10, as marked for vertically arranged panels. 
The horizontal solid and broken white lines mark the phases of the
primary and secondary eclipses. 
}
\label{fig14}
\end{center}
\end{figure}

\begin{figure}%[ht]
\begin{center}
\includegraphics[width=11.5cm]{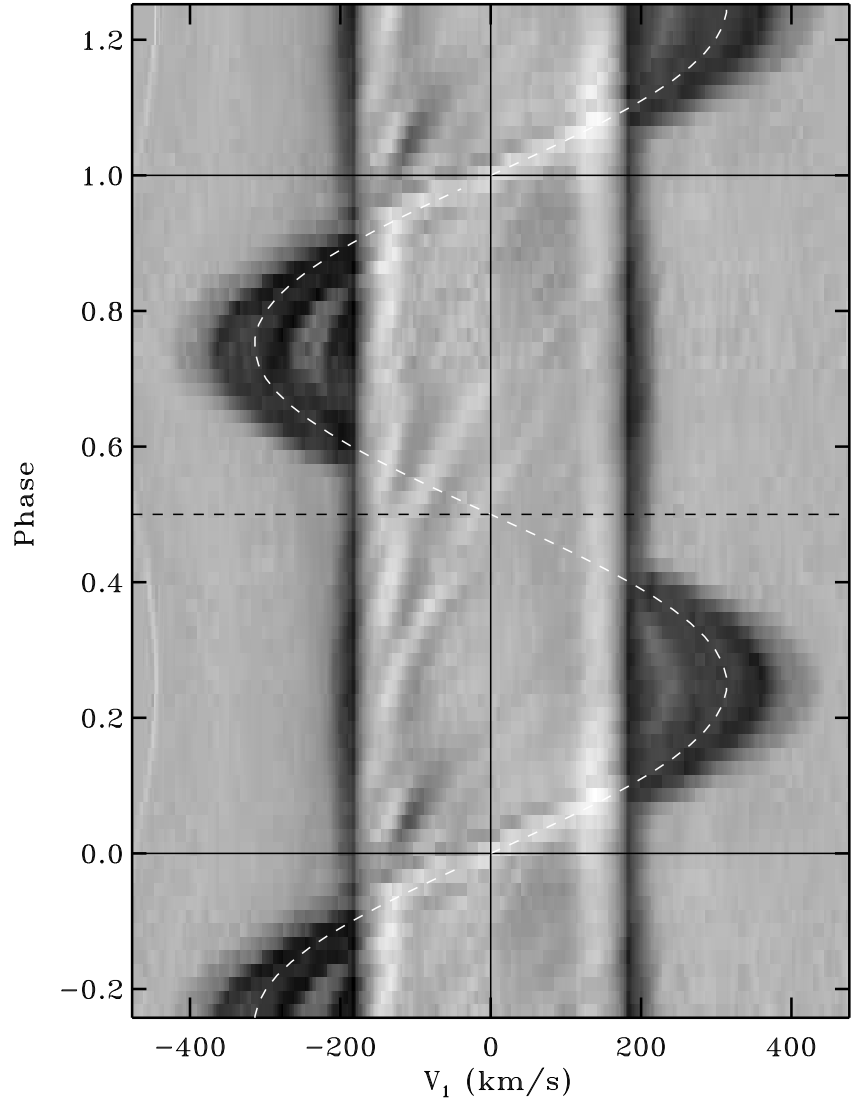}    
\caption{
\footnotesize 
The two-dimensional image of the AW~UMa system for all data
binned and averaged 
in phase and velocity. The horizontal axis gives velocities
in the system centered on the primary component.  
The sine curve gives the expected motion 
of the secondary for $K_1+K_2 = 315$ km~s$^{-1}$ (or $q = 0.099$). 
}
\label{fig15}
\end{center}
\end{figure}

\begin{figure}%[ht]
\begin{center}
\includegraphics[width=11.5cm]{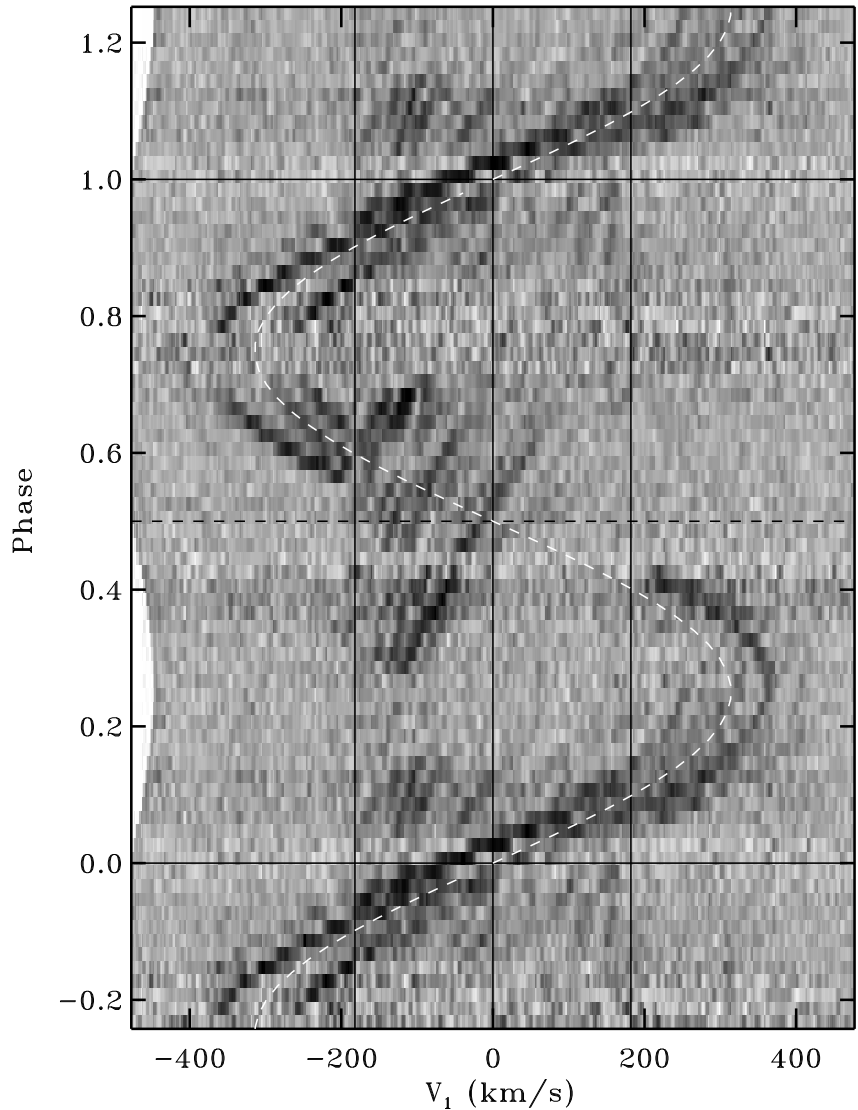}    
\caption{
\footnotesize 
The figure gives the standard deviations in the binned data as
shown in Figure~\ref{fig15}. Note that the most variable regions
are the peaks in the accretion structure around the secondary 
component and the spots on the primary component (best visible
for phases $0.35 < \phi < 0.65$). 
It is significant that the primary and -- in particular -- 
the pedestal around it do not seem to be variable in the 
night-to-night timescale.
}
\label{fig16}
\end{center}
\end{figure}

During the secondary star occultations (the secondary or shallower 
eclipses) when it disappears behind the large star, 
the velocity field of the secondary component is hardly modified at all. 
In Figures~\ref{fig8} -- \ref{fig10}, which
show deviations from the rotational profile, we can detect slight
indentations at the high-velocity edges of the secondary
profiles (around phases 0.55, 3.05, 5.05, 5.45, 5.55)
which suggest that the highest velocities are cut off. 
Otherwise, the whole system of radial velocities associated
with the secondary component is eclipsed as if it 
was a single, extended object. The phases of the eclipse
contacts are given in Table~\ref{tab_contact};
they were determined by eye from figures similar to 
Figures~\ref{fig8} -- \ref{fig10}. While
these determinations carry substantial uncertainty 
($\pm 0.01$ in phase) due to the presence of the pedestal, 
the phases of external contact 
(Contacts 1 and 4 in Table~\ref{tab_contact}) 
are definitely too large to be
identified with the real geometrical contact phases, 
but -- rather -- they reflect the large range of velocities  
in the accretion structure.

\placetable{tab_contact}          % Table 5: contact phases

The secondary star transits over the disk of the primary
component (the primary or deeper eclipses) appear to show 
contacts which are more similar to what is observed in the 
photometric light curves. 
In Figures~\ref{fig8} -- \ref{fig10}, 
the secondary component is visible as two or sometimes 
three bright ridges which continue the motion of the secondary 
as observed before and after the eclipses. It is
dark inside, but this mainly reflects the imperfect fits 
of the rotational profile during the transit phases.
Although the fits are indeed corrupted, the secondary 
does produce a very weak disturbance in the
primary profiles as can be seen in the last panel of
Figure~\ref{fig1}.  Unfortunately,
only one transit was observed exactly during the central 
eclipses phases (night 3, phase 5.0); however, the two other 
transits provided moderately well-defined phases of the external and 
internal contacts (Table~\ref{tab_contact}). 

The observations obtained in the center of occultation
at phase 5.0 (see Figures~\ref{fig1} and \ref{fig10}),
give the velocity extent of the secondary as ranging
between $-52$ and $+52$ km~s$^{-1}$, but with a faint extension
reaching on the positive side to $+90$ km~s$^{-1}$. 
Thus, the full radial-velocity extent of 104 km~s$^{-1}$
is larger than the {\it mean\/} peak 
separation during the orbital quadratures,  
81 km~s$^{-1}$ (Section~\ref{sec-vel}), although
on one occasion the separation reached 98 km~s$^{-1}$.
The amount of the light loss due to the secondary, within the above
velocity range is hard to estimate because the depression is very
shallow and comparable to depths of the inhomogeneities on the surface
of the primary component; very roughly it absorbed $0.05 \pm 0.02$
of the primary light. This crude estimate, when transformed to linear
dimensions gives the linear size of about $0.22 \pm 0.05$ relative to
the size of the primary component. Note, that for the Roche lobes with
$q = 0.1$, the ratio of the ``side'' or orbital-plane dimensions 
is about 0.35 -- 0.36 (weakly depending on the degree of fill-out). 

The inner contact phases during the transits when the secondary
touches the edges of the primary from inside
(Contacts 2 and 3) are distinctly 
asymmetric relative to the eclipse centre, with the mean values of
the phases $-0.061$ and $+0.034$, see Table~\ref{tab_contact}. While
the former is the same as the well determined 
angle of the photometric internal contact $0.062 \pm 0.005$ 
\citep{MD1972}, the exit contact phases are much smaller. 
The same asymmetry is visible in the
``light-curve'' of AW~UMa in Figure~\ref{fig2}.
Possibly, it may be explained by the shift in the direction 
of the mass flow close to the $L_1$ point which is quite prominent
in the model of \citet{Oka2002}.

\section{AW UMa as a binary system}
\label{binary}

\subsection{Spectroscopic orbital elements}    
\label{orbit}

With the orbital velocity amplitude of the primary component
$K_1 = 28.37 \pm 0.37$ km~s$^{-1}$ (Section~\ref{pri-orb})
and the sum of the orbital amplitudes determined from the 
orbital quadratures, 
$K_1 + K_2 = 314.2 \pm 3.2$ km~s$^{-1}$ (Section~\ref{sec-vel}),
we find the mass ratio $q_{\rm sp} = M_2/M_1 = 0.0991 \pm 0.0014$,
confirming the determination of \citet{PR2008}.
While this is the best value for the spectroscopic
mass ratio, as described previously (Section~\ref{sec-vel}), we adopt
a two times larger error so that 
$q_{\rm sp} = M_2/M_1 = 0.099 \pm 0.003$.
Even with this adjustment, our spectroscopic mass ratio
is several formal errors away from the photometric determinations 
which usually converge to a value within $q_{\rm ph} \simeq 0.07 - 0.08$,
but with individual errors usually determined very small, 
as small as 0.0005 (e.g.\ \citet{Wilson2008}). 
Using the measured $K_1$ and following our assumption on
$q_{\rm sp}$, the estimated orbital amplitude for the secondary is 
$K_2 = 286.6 \pm 12.4$ km~s$^{-1}$, 
the distance between the mass centers
$A \sin i = 2.73 \pm 0.11\, R_\odot$, and the masses
$M_1 \sin^3 i = 1.292 \pm 0.15\, M_\odot$ and 
$M_2 \sin^3 i = 0.128 \pm 0.016\, M_\odot$.

The orbital inclination of AW~UMa must be close to the edge-on
orientation; otherwise we would not see the long-lasting, total 
eclipses. We certainly cannot assume the orbital inclination angle of 
$i \simeq 78 - 80$ deg, as determined
from light-curve synthesis solutions published over several decades
investigation (starting with $80$ deg, \citet{MD1972}, with the
most recent $78$ deg, \citet{Wilson2008}), as they all assumed the
Lucy contact model. Thus, the above orbit-size and mass determinations
may be close to the actual values, but there will remain an additional
systematic uncertainty at a level of 1.5\% for the linear dimensions
and 5\% for the masses. 
% For $i=80$ degrees, $sin i = 0.985$ and $sin^3 i = 0.955$.

\subsection{The evolutionary state}
\label{evol}

AW~UMa appears to be a semi-detached binary showing mass transfer 
from the more massive to the less massive component. It is 
not a contact binary as envisaged in the contact binary model of Lucy:
The complex internal velocities and obvious accretion processes on or
around the low-mass secondary component invalidate the contact model
for this binary. 

In terms of the evolutionary status of the binary, the new data 
fully support the models developed by \citet{Step2006} and 
\citet{BeP2007}. The secondary is probably the helium core of
a former more massive component which already evolved past the mass
exchange; now it is the formerly less-massive component which 
expands and sends matter to the small star.
In terms of its kinematics and metallicity, AW~UMa appears to belong 
to the local solar population \citep{Rci2013a}: the spatial velocity
components 
$[U, V, W] = [12.5\pm0.5, -55.5\pm4.1, -9.5\pm0.5]$ km~s$^{-1}$
and the metallicity
$[M/H] = -0.01 \pm 0.08$ are typical for an age of 3 -- 5 few Gyrs.  
This age would be long enough to produce a system already in the
stage of the former secondary expanding beyond its Roche limit. 

With its systematically shortening orbital period,
AW~UMa appears to be evolving towards an eventual merger. 
Thus, in the time scale of 
$(d\ln P/dt)^{-1} \simeq 2 \times 10^6$ years, 
AW~UMa will show an eruption and coalescence
in the same way as was observed and analyzed recently by
V1309~Sco \citep{Tyl2011,Step2011,Pej2014}.

\section{Conclusions and directions for further research}
\label{further}

Except for the important conclusion that AW~UMa is not a 
contact binary, but a semi-detached system, the
present work has produced more questions than answers...
First of all, the above conclusion leads to a major disagreement 
with the excellent explanation of its light curve by the 
Lucy contact-binary model. 
Why do the light curve synthesis models, based on the contact model,
give such perfect reproductions
of the AW~UMa light curves and yet spectroscopy gives an entirely
different, complex picture with many asymmetries and variations
within a semi-detached binary? Is something missing in the current
analysis?

The phase dependence of the integrated Broadening Functions
indicates that most of the photometric variations are produced by the
same material; one can even restore the light curve assuming 
the constancy of the primary component (the remaining difference
is 3\% for the full 19\% photometric modulation). Thus, we seem to see the 
same matter which causes light variations and produces spectra.
Of course, it is true that symmetric mass distribution 
(along the line of mass-centers) may still contain (and hide)
asymmetric motions and that the BF method is only sensitive to
velocities. But are the discrepancies not too large? This major
disagreement requires further study because rejection of the contact
model should not be done too hastily: It is a conceptually
simple and consistent model which apparently has worked well since
its inception in 1968 \citep{Lucy1968a,Lucy1968b}. But the model still
requires rigorous tests on its applicability; the observations such as
those presented here for other W~UMa-type binaries would be most useful.
After all, the current study is the first in-depth, high-resolution,
high signal-to-noise, spectroscopic study of a W~UMa binary. 
Because of the rapid variations, high-resolution spectroscopic studies 
require allocation of time on large telescopes, at least of the
4-meter class, together with efficient spectrographs. 
The most obvious target for such
a study would be $\epsilon$~CrA, the brightest W~UMa binary in the sky
($V=4.8$). Its period is slightly longer (0.591 d) and the spectral type
later (F3), but it can be considered almost a twin of AW~UMa in 
that it also shows total eclipses and its mass ratio is small.
The spectroscopic study of \citet{GD1993} led to determination
of masses, both slightly larger than that of AW~UMa, but it was
conducted at too low a resolution, $R \simeq 10,000$, to be able to 
reveal complexities as extensive as those observed for AW~UMa. 
It is interesting that the spectroscopic mass ratio 
was also found for $\epsilon$~CrA to be larger than the photometric one.

While the primary component of AW~UMa seems to be a simple, fast rotating,
non-variable F2V star, it shows unexpected features when analyzed 
in greater detail: The current work has confirmed the existence
of the ``pedestal'' of large rotational/orbital velocities
around the primary and of numerous inhomogeneities on its surface. 
\citet{Step2009} has already given an explanation of the pedestal in
terms of the extensive equatorial-plane flow which results from 
the mass-exchange process in the binary; the study was partly
inspired by the previous analysis of AW~UMa by \citet{PR2008}.
The pedestal, which reaches a total intensity of about 4\%
of the primary star, is surprisingly constant between the
individual orbital cycles. In addition, it shows the double-cosine 
variability which is in phase with the photometric ellipticity effect.  
The pedestal seems to be the only phase-dependent feature which shows 
a full symmetry along the axis joining
the stars. The inhomogeneities on the primary 
were suspected in the data of \citet{PR2008}, but only
now, with continuous temporal coverage on three consecutive night
have they been seen in full detail. The dense network of
the ``ripples'' has no immediate explanation. 
The ripples share the photospheric rotation of the primary yet they seem to 
extend into the pedestal; their amplitudes grow toward lower
spatial frequencies. The ``spots'' are 
equally mysterious: Why do they appear only at negative velocities? 
What are they? Unfortunately, our analysis leaves  room for
interpretation as the dark notches in the BF's may equally be caused 
by genuine photospheric dark spots as well as locally 
shallower spectral lines, either due to higher temperatures or to
filling-in by emissions. The very slow rotation of the spots
suggests their participation in a strongly differential rotation,
perhaps by $\simeq$~50\% slower than the star itself; why then don't 
we see any discrepancies in the rotational-profile fits which imply
a consistent value of $V \sin i = 181$ km~s$^{-1}$? 

The secondary component shows complex accretion phenomena. It 
appears as a system of bands or strands which change in time so
it is hard to tell what the actual shape of the star is.
It must be very small, as its transits in front of the 
primary produce very small line-profile effects.
The local centre of velocities, which we identify as the 
secondary component, moves as for the binary mass ratio
$q_{\rm sp} = M_2/M_1 = 0.099 \pm 0.003$. This is a result
many formal errors away from the previous, photometric, 
very consistent determinations. Which value is the correct one? Is
this because the light curves have a relatively low information 
content and we only now have a proper dynamical determination? The
light curves are indeed featureless except for the well defined 
inner eclipse contact, whose phase very strongly constraints 
the synthesis solutions, particularly the value of $q_{\rm ph}$. While
we do see inner eclipse contacts also in velocities, they appear to be
different than the photometric ones and variable from eclipse to eclipse. 
The mean velocities of the secondary component permit determination of 
minimum masses for the components of AW~UMa: 
$M_1 \ge 1.29 \pm 0.15\, M_\odot$ and 
$M_2 \ge 0.128 \pm 0.016\, M_\odot$ which are most likely very close
to the actual ones because the orbit is seen close to the
edge-on orientation. 
 
As was said above, this study has led to more questions than answers.
While the evolutionary and physical state of AW~UMa is  
moderately simple,
the many new features discovered in the detailed view of AW~UMa require
further investigation. Further studies based on the same excellent
CFHT spectra are planned as the followup of the current work.

\begin{acknowledgements}

The author would like to express his most sincere thanks
to several individuals who helped him during this work either directly
or through extensive discussions and comments:
Jano Budaj, Jean-Francois Donati, Nadine Manset, Theo Pribulla, 
Micha\l\  R\'{o}\.{z}yczka, 
Kazik St\c{e}pie\'n, Bob Wilson, Walter Van Hamme, Gemma Whittaker.

Thanks are due to the reviewer of the first version of the manuscript
for the very pointed and constructive comments and suggestions.

The research of the author has been supported by the
Natural Sciences and Engineering Research Council of
Canada.  This research made use of the SIMBAD database, 
operated at the CDS, Strasbourg, France and of NASA's 
Astrophysics Data System (ADS).

\end{acknowledgements}

\newpage
%============================================================================
% Table 1: observations

\begin{deluxetable}{cccc}

\tabletypesize{\footnotesize}    % 10 pts
\tablewidth{0pt}
\tablecaption{Observations of AW~UMa \label{tab_obs}}
\tablenum{1}

% night  HJD   phase interval  no. of spectra

\tablehead{\colhead{Date 2011} & \colhead{Range} &
   \colhead{$\phi$} & \colhead{$N$} 
}

\startdata
 11 March & 55,631.850 -- 55,632.146 & 0.456 -- 1.131 & 152 \\
 12 March & 55,632.746 -- 55,633.147 & 2.499 -- 3.412 & 200 \\
 13 March & 55,633.781 -- 55,634.141 & 4.858 -- 5.567 & 232
\enddata

\tablecomments{The time range for mid-exposures is
expressed as $HJD - 2,400,000$; $\phi$ gives the range
in phase computed from the assumed initial epoch $HJD = 2,455,631.6498$ 
and $P = 0.43872420$ day; $N$ is the number of spectra 
obtained on a given night.}

\end{deluxetable}
%---------------------------------------------------------------------------
% Table 2: ASCII BF's

\begin{deluxetable}{rrrrrrrr}

\tabletypesize{\footnotesize}          % 10 pts
\tablewidth{0pt}
\tablecaption{Broadening Functions \label{tab_BFs}}
\tablenum{2}

\tablehead{ &  &  &  &  &  &  &  
%\tablehead{\colhead{$T$}  &  &  &  &  &  &  &  
}

\startdata
       & $-4950$ & $-4932$ & $-4914$ & $-4896$ & $-4878$ & $-4860$ & ... \\ 
 18497 & $ -700$ & $ -400$ & $  -49$ & $  128$ & $  141$ & $  106$ & ... \\
 18512 & $ -959$ & $ -716$ & $ -362$ & $  -87$ & $   76$ & $  135$ & ... \\
 18527 & $ -821$ & $ -643$ & $ -348$ & $ -110$ & $    8$ & $   15$ & ... \\ 
 ...   &         &         &         &         &         &         &  
\enddata

\tablecomments{
This table is available in the on-line version only.\\
For efficient transfer, the ASCII file has
all numbers scaled and converted to integers. 
The first row gives 551 velocities in the heliocentric system, 
expressed in km~s$^{-1}$ and multiplied by 10 (format: 6x,551i6). 
The subsequent 584 rows (format: 552i6) give, in the first position, 
the time 
%as $T = 
$HJD-2\,455\,630$ multiplied by $10^4$,  and then 
551 values of the Broadening Function multiplied by $10^6$.
Thus, the velocities in the first row are: 
$-495.0$, $-493.2$, $-491.4$, $-489.6$, $-487.8$, $-486.0$, ... km~s$^{-1}$
and the corresponding BF values at $T = 1.8497$ are: 
$-0.000700$, $-0.000400$, $-0.000049$, $0.000128$, $0.000141$, $0.000106$. 
}

\end{deluxetable}
%---------------------------------------------------------------------------

% Table 3:    ASCII radial velocities  

\begin{deluxetable}{ccccc}

\tabletypesize{\footnotesize}          % 10 pts
\tablewidth{0pt}
\tablecaption{Radial velocities of the primary component\label{tab_RVs}}
\tablenum{3}

\tablehead{ \colhead{$T$}  & \colhead{$\phi$} &
    \colhead{~~~$V_1^{45}$}  & \colhead{~~~$V_1^{50}$}  & 
    \colhead{~~~$V_1^{55}$}  }

\startdata
 1.8497 & 0.4556 & $-22.40$ & $-22.20$ & $-21.19$ \\
 1.8512 & 0.4590 & $-22.60$ & $-23.30$ & $-23.20$ \\
 1.8527 & 0.4624 & $-22.50$ & $-22.96$ & $-23.06$ \\
 1.8542 & 0.4658 & $-22.69$ & $-23.33$ & $-23.00$ \\
 1.8559 & 0.4697 & $-20.46$ & $-21.02$ & $-21.27$ \\
 ...   &         &          &          &         
\enddata

\tablecomments{
This table is available in the on-line version only.\\
The mid-exposure times are given as
$T = HJD-2\,455\,630$, while the phases are computed with the assumed
initial epoch $HJD = 2,455,631.6498$ and the period 0.43872420 day.
RV measurements for all observations
are listed, but only those within the fractional phase interval 0.25 -- 0.75
were used for the orbital solutions. The primary star velocities 
(in km~s$^{-1}$) are given in the three columns labeled 
$V_1^{45}$, $V_1^{50}$ and $V_1^{55}$ corresponding to the low
cut-off to the rotational profile fit at 0.45, 0.50 and 0.55
of the BF maximum.
}

\end{deluxetable}
%---------------------------------------------------------------------------

% Table 4:    orbital elements 

\begin{deluxetable}{cccc}

\tabletypesize{\footnotesize}    % 10 pts
\tablewidth{0pt}
\tablecaption{Orbital solutions for the primary component \label{tab_orb}}
\tablenum{4}

\tablehead{\colhead{Cut-off} & 
   \colhead{$V_0$} & \colhead{$K_1$} & \colhead{$\sigma_1$}
}

\startdata
0.45  &  $-15.78 \pm 0.06$  &  $28.67 \pm 0.085$ & 0.99 \\
0.50  &  $-15.90 \pm 0.07$  &  $28.33 \pm 0.102$ & 1.20 \\
0.55  &  $-16.01 \pm 0.08$  &  $28.12 \pm 0.115$ & 1.34 \\
      &                     &                    &      \\
Mean  &  $-15.90 \pm 0.12$  &  $28.37 \pm 0.37$  &      
\enddata

\tablecomments{The rotational profile fits were done above the 
Cut-off level expressed relative to the BF maximum. 
$\sigma_1$ is the fit error for one observation. 
All velocities are in km~s$^{-1}$.
The errors for the Mean values represent the $rms$ 
scatter of the determinations, not the formal Gaussian errors 
of the mean (i.e.\ divided by $\sqrt(3)$), 
because the deviations are systematic.
}

\end{deluxetable}
%---------------------------------------------------------------------------
% Table 5: contact phases

\begin{deluxetable}{ccccc}

\tabletypesize{\footnotesize}    % 10 pts
\tablewidth{0pt}
\tablecaption{Eclipse contact phases \label{tab_contact}}
\tablenum{5}

% night  c1   c2   c3   c4

\tablehead{\colhead{2011} & 
   \colhead{Contact 1} & \colhead{Contact 2} &
   \colhead{Contact 3} & \colhead{Contact 4}
}

\startdata
\sidehead{Transit} 
% 11 March & 0.850 & 0.938 & 1.032 & 1.128 \\
 11 March & $-0.150 + 1$ & $-0.062 + 1$ & +0.032 + 1 & +0.128 + 1 \\
% 12 March & 2.871 & 2.940 & 3.034 & 3.128 \\
 12 March & $-0.129 + 3$ & $-0.060 + 3$ & +0.034 + 3 & +0.128 + 3 \\
% 13 March & 4.860 & 4.940 & 5.037 & 5.117 \\
 13 March & $-0.140 + 5$ & $-0.060 + 5$ & +0.037 + 5 & +0.117 + 5 \\
\sidehead{Occultation}
% 11 March &  \nodata & \nodata & 0.544 & (0.610) \\
 11 March &  \nodata & \nodata & +0.044 + 0.5 & $(+0.110)$ + 0.5 \\
% 12 March & (3.340) & \nodata  & (2.555) & (2.633) \\
 12 March & $(-0.160)$ + 3.5 & \nodata  & $(+0.055)$ + 2.5 & $(+0.133)$ + 2.5 \\
% 13 March & (5.360) & 5.436 & 5.554 & \nodata  \\	
 13 March & $(-0.140)$ + 5.5 & $-0.064 + 5.5$ & +0.054 + 5.5 & \nodata  \\	
\enddata

\tablecomments{Phases of contacts are given from the first outer 
Contact~1, through the two inner Contacts~2 and 3 to the
outer Contact~4. The format for the phases, with separately expressed
integer parts, is used to emphasize 
the phase difference relative to the eclipse center yet retain
information on when the eclipse was observed within the run.
Errors in phase are about 0.005 for transits and about 0.01
for occultations; values less accurate are in parentheses.}

\end{deluxetable}
%---------------------------------------------------------------------------

%===========================================================================
% figures

%               4 phases        fig1.eps 

% ---------------------------------------------------------------
%               strengths    fig2.eps 
%----------------------------------------------------------------

%               Vsini          fig3.eps 
%----------------------------------------------------------------
%               orbit             fig4.eps 
%----------------------------------------------------------------
%               deviations in RV          fig5.eps 
%----------------------------------------------------------------
%               4 devs          fig6.eps 
%----------------------------------------------------------------

%               pedestal        fig7.eps   
%----------------------------------------------------------------

%               2-D image        fig8.eps 
%----------------------------------------------------------------

%               2-D image        fig9.eps  
%----------------------------------------------------------------

%               2-D image        fig10.eps  
%----------------------------------------------------------------

%               ripples        fig11.eps   
%----------------------------------------------------------------
%               spots        fig12.eps   

% to avoid 'Too many unprocessed floats'
\clearpage
%----------------------------------------------------------------
%               spots        fig13.eps 
%----------------------------------------------------------------
%               2D secondary        fig14.eps 

%----------------------------------------------------------------
%               2D average secondary      fig15.eps   

%----------------------------------------------------------------
%               2D average secondary      fig16.eps 

\end{document}